\documentclass[a4paper,11pt]{article}

\usepackage{jcappub} 

\usepackage[T1]{fontenc} 
\usepackage{xcolor}
\usepackage{bm}
\usepackage{appendix}
\usepackage{enumerate}

\title{Prospects for Constraining the Yukawa Gravity with Pulsars around
Sagittarius A*}

\author[a,b]{Yiming Dong,}
\author[b,c,1]{Lijing Shao,\note{Corresponding author.}}
\author[a,b]{Zexin Hu,}
\author[c]{Xueli Miao}
\author[a,b]{and Ziming Wang}

\affiliation[a]{Department of Astronomy, School of Physics, Peking University,
Beijing 100871, China}
\affiliation[b]{Kavli Institute for Astronomy and Astrophysics, Peking
University, Beijing 100871, China}
\affiliation[c]{National Astronomical Observatories, Chinese Academy of
Sciences, Beijing 100012, China}

\emailAdd{lshao@pku.edu.cn}

\abstract{The discovery of radio pulsars (PSRs) around the supermassive black
hole (SMBH) in our Galactic Center (GC), Sagittarius A* (Sgr A*), will have
significant implications for tests of gravity. In this paper, we predict
restrictions on the parameters of the Yukawa gravity by timing a pulsar around
Sgr A*  with a variety of orbital parameters.  Based on a realistic timing
accuracy of the times of arrival (TOAs), $\sigma_{\rm TOA}=100\,\mu{\rm s}$, and
using a number of 960 TOAs in a 20-yr observation, our numerical simulations
show that the PSR-SMBH system will improve current tests of the Yukawa gravity
when the range of the Yukawa interaction varies between $10^{1}$--$10^{4}\,{\rm
AU}$, and it can limit the graviton mass to be $m_g \lesssim 10^{-24}\,{\rm
eV}/c^2$.}

\begin{document}

\maketitle
\flushbottom

\section{Introduction}
\label{Sec1}

General relativity (GR) has stood up to a variety of tests since its proposal by
Albert Einstein in 1915~\cite{Will:2014kxa,Berti:2015itd}. However, the
phenomena of dark matter~\cite{Bertone:2016nfn} and dark
energy~\cite{Debono:2016vkp} have posed a great challenge to GR and inspired
proposals for a series of modified gravity
theories~\cite{Moffat:2004nw,Saridakis:2021vue}.  Theories of  massive
gravity~\cite{Hinterbichler:2011tt}, which add a mass term to the graviton, have
aroused wide interest.  Some earlier theoretical difficulties were eventually
overcome, for example, in the resolution to the van Dam-Veltman-Zakharov (vDVZ)
discontinuity~\cite{vanDam:1970vg, Zakharov:1970cc} via the Vainshtein
mechanism~\cite{Vainshtein:1972sx}. There are many massive gravity
theories~\cite{deRham:2014zqa}, and a popular parametrization is called the
Yukawa potential theory. In this theory, the gravitational potential is
Yukawa-type suppressed~\cite{Hinterbichler:2011tt, Borka:2013dba,
Zakharov:2016lzv, Zakharov:2018cbj}.  For a point mass source, the gravitational
potential is expressed as,
\begin{equation}
    \phi(r)=-\frac{GM}{(1+\alpha)r}\left( 1+\alpha {\rm
    e}^{-\frac{r}{\Lambda}}\right),
\label{eq:Yukawa_potential}
\end{equation}
where $G$ is the gravitational constant, $M$ is the mass of the central object,
$r$ is the distance to central object, $\alpha$ and $\Lambda$ represent the
strength and the range of the Yukawa interaction respectively. According to
Eq.~(\ref{eq:Yukawa_potential}), when the length scale $r$ of a system is much
less than the range of the Yukawa interaction $\Lambda$, the Newtonian potential
$\phi_{\rm N}(r) \propto - GM/r$ is recovered so, augmented with post-Newtonian
(PN) corrections, it passes the stringent gravity tests from the Solar system
when $\Lambda \gg 1$\,AU~\cite{Talmadge:1988qz,Tsai:2021irw}. When $r$ is
comparable to $\Lambda$, the Yukawa gravity will show clear deviations from the
Newtonian inverse-square law.  If $r \gg \Lambda$, a Newtonian-like potential is
recovered but with a different gravitational constant, $G_{*}=G/(1+\alpha)$. The
change of gravitational constant has been successfully used to account for the
observational rotation curves of spiral galaxies in a special case of
$\alpha=1/3$~\cite{Cardone:2011ze}. 

There are generally two ways to constrain the parameters of the Yukawa gravity.
The first way is to obtain the constraints on the strength of the Yukawa
interaction $\alpha$ as a function of the range of the Yukawa interaction
$\Lambda$, which is used to search for new kinds of interactions, the so-called
fifth force~\cite{Talmadge:1988qz,Fischbach:1992fa}.  Thus the Yukawa gravity
can be a probe to seek for new kinds of interactions and fundamental fields,
which could be candidates for dark matter and dark
energy~\cite{Raffelt:1999tx,Adelberger:2006dh,Tsai:2021irw}.  Due to the
characteristics of the Yukawa potential~(\ref{eq:Yukawa_potential}), the optimal
limit on $\alpha$ at a certain $\Lambda$ is usually obtained when the system
scale $r$ is similar to $\Lambda$.  Therefore, to fully explore the Yukawa
gravity, it is necessary to carry out tests at a wide range of length scales.
Tests have been conducted at various length scales, such as torsion oscillators
at the laboratory scale~\cite{Niebauer:1987ua,Hoyle:2000cv,Adelberger:2003zx},
LAser GEOdynamic Satellite (LAGEOS) at a length scale of the
Earth~\cite{Peron:2014pba}, and Lunar Laser Ranging (LLR)~\cite{Williams:2005rv,
Muller:2005sr} at the Earth-Moon system scale, and planetary
orbits~\cite{Talmadge:1988qz, KONOPLIV2011401, Hees:2014kta, Li:2014hya,
Tsai:2021irw, Benisty:2022txp} at the Solar system scale. Stringent limits on $\alpha$ have been
acquired in the fifth-force framework except at extremely small and extremely
large $\Lambda$'s.  Sagittarius A* (Sgr A*), a supermassive black hole (SMBH)
with a mass of $\sim 4\times10^{6}\,M_{\odot}$, exists in our Galactic Center
(GC)~\cite{Gillessen:2008qv, EventHorizonTelescope:2022xnr}. If we accurately
measure the motion of objects around the central gravitational source, we can
test the Yukawa gravity in the strong-field situation. Besides, the orbital
semi-major axes of objects around Sgr A* can reach up to $10^{3}\,{\rm AU}$,
which are much larger than the length scales of what have been tested before. It
has the potential to fill in the blank in tests of the Yukawa gravity at a large
$\Lambda$. Efforts to constrain the fifth force by measuring the orbits of stars
around Sgr A* have come to fruition~\cite{Hees:2017aal}.  However, compared to
the limits on $\alpha$ from the Solar system, the limits are relatively weak
because of the relatively limited precision in imaging stars' orbits.

The second way to constrain the parameters of the Yukawa gravity is to limit the
range of the Yukawa interaction $\Lambda$, which effectively corresponds to
limiting the graviton mass $m_g$.  The range of the Yukawa interaction $\Lambda$
can be regarded as the Compton wavelength of the massive
graviton~\cite{Hees:2017aal, Bernus:2019rgl}.  If an object is influenced by the
Yukawa gravity instead of the Newtonian one, its orbit will show deviations from
the Keplerian orbit, for example, with an additional contribution in the
periastron advance rate~\cite{Will:2018gku}.  Based on the analysis of bright
stars' trajectories around Sgr A*, there are some bounds on the graviton mass
provided in Refs.~\cite{Borka:2013dba, Zakharov:2016lzv, Zakharov:2018cbj,
Jovanovic:2021hrz}. Zakharov et~al.~\cite{Zakharov:2018cbj} predicted the
upper limits on $m_{g}$ from stars around Sgr~A*; they showed the bound will
reach $\sim 5\times10^{-23}\,{\rm eV}/c^{2}$ from a star with $P_{{b}}=50\,{\rm
yr}$ and a small eccentricity. Other methods have been suggested to bound the
graviton mass~\cite{deRham:2016nuf}, such as using the orbital motion of planets
in the Solar system~\cite{Will:2018gku, Bernus_2020, Iorio:2007gq} and binary
pulsars~\cite{Finn:2001qi, Miao:2019nhf, deRham:2012fw, Shao:2020fka}. The bound
on $m_{g}$ from the Solar system can reach $3.62\times10^{-23}\,{\rm eV}/c^{2}$
at 99.7\%~C.L. through ephemeris INPOP19~\cite{Bernus_2020}.  The modified
dispersion relation of gravitational waves (GWs) provide a new opportunity to
limit the graviton mass~\cite{Will:1997bb}.  Due to the modified dispersion
relation, low-frequency GWs will travel slower than high-frequency ones,
resulting in deformed gravitational waveforms, providing an avenue to set limits
on the graviton mass with matched filtering
techniques~\cite{LIGOScientific:2016lio}. The latest bound from the
LIGO-Virgo-KAGRA Collaboration on $m_{g}$ is $1.27\times10^{-23}\,{\rm
eV}/c^{2}$ at 90\%~C.L. from the GW transient catalog
GWTC-3~\cite{LIGOScientific:2021sio}. Similarly, we can perform tests with
multi-messenger detections utilizing the time or phase differences between GWs
and electromagnetic signals~\cite{Piorkowska-Kurpas:2022xmb}.

Pulsars (PSRs), identified as rotating neutron stars with highly stable
rotational periods, provide a powerful method to test GR and other fundamental
theories~\cite{Taylor:1992kea, Shao:2016ezh, Wex:2014nva, Kramer:2016kwa,
Miao:2020wph, Shao:2022izp, Shao:2022koz}. Pulsar timing technology is the
common method for processing the times of arrival (TOAs) of radio signals from
pulsars. For binary pulsars, it involves precisely measuring the TOAs of pulses
and modeling them appropriately to get the information on the orbital motion of
the pulsar, which in turn helps us test the gravitational dynamics that underpin
the orbital motion~\cite{Taylor:1992kea}.  The first discovered binary pulsar,
PSR B1913+16, indirectly proved the existence of GWs by measuring the binary
orbital period decay rate~\cite{Hulse:1974eb, Taylor:1979zz}. The discovery of
the Double Pulsar, PSR J0737$-$3039A/B, and the measurement of its many
post-Keplerian parameters, including the periastron advance and the parameters
of the Shapiro delay, enabled us to test theories of gravity with unprecedented
precision~\cite{Kramer:2006nb}. It provides a versatile testbed of GR and some
first measurements of the higher-order relativistic
effects~\cite{Kramer:2021jcw}.  Pulsars also played an important role in
limiting the graviton mass~\cite{Finn:2001qi, Miao:2019nhf, deRham:2012fw,
Shao:2020fka}.  In massive gravity, GWs will take away extra energy of a binary
system, which results in an accelerated orbital period decay rate.  So one can
measure the orbital period decay rate through binary pulsars to obtain bounds on
the graviton mass. As we will show, it is also an excellent option to use binary
pulsar systems to test the Yukawa gravity in its conservative dynamics.

Combining the above insights, by timing a pulsar around Sgr A*, one can make
better use of the PSR-SMBH system in gravity tests. The timing results could
provide an unparalleled opportunity to test the Yukawa gravity. Stars around Sgr
A* have been discovered and observed with high cadence~\cite{Gillessen:2008qv,
2017ApJ...837...30G}, while no suitable pulsar, which can be used as a
gravitational laboratory, has been found around Sgr A* yet~\cite{Liu:2021ziv,
Torne:2021yad, Suresh:2022vmf}. However, theories predict the existence of such
pulsars~\cite{Cordes:1996bt, Wharton:2011dv, Zhang:2014kva, Bower:2018mta,
2019BAAS...51c.438B}, and finding them is recognized as an important project
for, e.g.\ the Square Kilometre Array (SKA) and the next-generation Very Large
Array (ngVLA).  If a pulsar around Sgr A* is detected by the SKA or ngVLA, we
can expect to obtain more accurate measurements of the SMBH system at our GC and
perform more stringent gravity tests based on the pulsar
timing~\cite{Liu:2011ae, Shao:2014wja, Weltman:2018zrl}. 

In this paper, using dedicated numerical simulations we predict constraints on
the theory parameters of the Yukawa potential (\ref{eq:Yukawa_potential}) if a
PSR-SMBH system is detected. In Section~\ref{Sec2}, we introduce our numerical
integration of pulsars' orbits and the pulsar timing simulation of TOAs, as well
as two methods of parameter estimation on bounding the Yukawa gravity. In
Section \ref{Sec3}, we explore different PSR-SMBH systems, and report the
results from simulations on individual limits of the strength and range of the
Yukawa gravity, and also of their simultaneous bounds. Section \ref{Sec4}
concludes our study and presents a brief outlook.

\section{Setup of Pulsar Timing Simulation}
\label{Sec2}

We present the setup of pulsar timing simulation in this section. First, we
introduce a theoretical timing model and generate the TOA data from the model
with numerical integration of pulsar orbits. Then using our timing model we fit
the data with a realistic Gaussian noise  and get the residuals between the
observed data and what is predicted by the timing model.  Finally, with the
parameter estimation method, we get the measurement uncertainties of the
parameters.  The measurement uncertainties of the theory parameters, $\alpha$
and $\Lambda$, demonstrate the PSR-SMBH system's ability to constrain the Yukawa
gravity.
 
A timing model needs to provide theoretical TOAs by considering various
time-delay effects.  In our binary timing model, we need to accurately describe
a pulsar's orbital motion in a PSR-SMBH system.  We invoke the first PN
equations of motion with a modified Yukawa term for the orbital dynamics of a
pulsar around Sgr A*. The acceleration of the pulsar reads~\cite{Einstein:1938yz,Gravity},
\begin{equation}
\begin{aligned}
    \bm{a}=&-\frac{GM}{r^{3}}\bm{r}+\frac{\alpha}{1+\alpha}\frac{GM}{r^{2}}
    \left[ \frac{1}{r}-\left( \frac{1}{r}+\frac{1}{\Lambda}\right){\rm
    e}^{-\frac{r}{\Lambda}}\right]\bm{r} \\
    & + \frac{GM}{c^{2}r^{3}}\left[ \left(\frac{4GM}{r}-\dot{\bm{r}} \cdot
    \dot{\bm{r}} \right) \bm{r} +4\left(\bm{r} \cdot \dot{\bm{r}} \right)
    \dot{\bm{r}} \right]\,,
\end{aligned}
\label{Eq_acceleration}
\end{equation}
where $M$ is the mass of Sgr A*, $r$ is the distance to Sgr A*, $c$ is the speed
of light in vacuum, $\bm{r}$ and $\dot{\bm{r}}$ are the relative position vector
and velocity vector of a pulsar from the binary's barycenter, respectively. We
only consider the leading-order effect for the Yukawa gravity and have used the
first PN order terms from GR. On the right-hand side of Eq.~(\ref{Eq_acceleration}), the first term is just the acceleration in Newtonian gravity, the second term is the leading-order correction from the Yukawa gravity, and the last term is the first PN correction from GR which has been introduced in detail in Refs.~\cite{Einstein:1938yz,Gravity}. For simplicity, we also ignore the interactions
of other objects around Sgr A*, which holds true for relatively tight
orbits~\cite{Liu:2011ae}.  Due to the extreme mass ratio $m_{p}/M < 10^{-6}$
with $m_{p}$ being the pulsar mass, we treat the pulsar as a test particle.
Starting from Eq.~(\ref{Eq_acceleration}), we use the \texttt{scipy} package to
calculate the orbital integral with a relative error of $10^{-13}$.

For the timing model, we only consider the time delays from the effects of the
pulsar's orbital motion~\cite{BT},
\begin{equation}
    t_{\rm arr} = T_{\rm em}+\Delta_{{\rm R}}+\Delta_{{\rm S}}+\Delta_{{\rm E}},
\end{equation}
where $t_{\rm arr}$ is the arrival time of pulses at the Solar system barycenter
and $T_{\rm em}$ is the proper time of pulses when they are emitted;
$\Delta_{\rm{R}}$ is the R$\rm{\Ddot{o}}$mer delay, $\Delta_{\rm{S}}$ is the
Shapiro delay and $\Delta_{\rm{E}}$ is the Einstein delay~\cite{BT,
Damour:1991rd}.  We consider the lowest order of these three delay
effects~\cite{BT, Damour:1991rd},
\begin{align}
    \Delta_{\rm{R}}&=-\frac{1}{c}\bm{n}\cdot\bm{r}\,,\\
    \Delta_{{\rm S}} &= \frac{2GM}{c^{3}}
    \log\left(\frac{2r_{\rm{s}}}{r+\bm{r}\cdot\bm{n}}\right)\,,\\
    \Delta_{{\rm E}}&=t_{\rm em} - T_{\rm em}\,,\\
    \frac{{\rm d} T}{{\rm d} t}&=1+\frac{GM}{c^{2}r}-\frac{1}{2}\frac{v^{2}}{c^{2}}\,,
    \label{Eq_Einstein_delay}
\end{align}
where $\bm{n}$ is the direction vector from the barycenter of binary to the
barycenter of the Solar system, $r_{\rm{s}}$ is the distance to barycenter of
the Solar system from the pulsar, $v=|\dot{\bm{r}}|$, $T$ and $t$ are the pulsar
proper time and the coordinate time respectively. The subscript ``$\rm{em}$''
means the time when pulses are emitted from the pulsar, which is related to the
pulsar's proper rotation number $N$ via \citep{Damour:1991rd},
\begin{equation}
    N(T)=N_{0}+\nu T+\frac{1}{2} \dot{\nu} T^{2}+\frac{1}{6}\ddot{\nu} T^{3}+ \cdots \,,
    \label{eq:numbernu}
\end{equation}
where $N_{0}$ is the initial rotation number and $\nu$ is the spin frequency of
pulsar. In our work, we ignore $\ddot{\nu}$ and higher order terms, as well as
the correction from Yukawa gravity in Eq.~(\ref{Eq_Einstein_delay}).  The effect
of the Shapiro delay is usually detectable only when the orbital plane of a
binary pulsar is nearly edge-on. However, for a PSR-SMBH binary system with a
small orbital inclination, the Shapiro delay is still evident due to the large
mass of the SMBH. In addition, the spin of Sgr~A* introduces significant effects
in general~\cite{Wex:1998wt, Liu:2011ae}. A complete timing model for PSR-SMBH
systems is still under development in the community, but the simple setting here
is sufficient for our purposes to obtain leading-order constraints on the Yukawa
gravity.

The parameters of the timing model include the mass of Sgr A* $M$, the spin
frequency of pulsar $\nu$, the time derivative of spin frequency $\dot{\nu}$,
the initial rotation number $N_{0}$, the orbital eccentricity $e$, the orbital
period $P_{b}$, the orbital inclination $i$, the longitude of periastron
$\omega$, the longitude of initial phase $\varphi_{0}$, the strength of the
Yukawa interaction $\alpha$, and the range of the Yukawa interaction $\Lambda$.
Collectively, we have the parameter set,
\begin{equation}
    \bm{\Theta}=\{M,\ \nu,\ \dot{\nu}, \ N_{0},\ e,\ P_{{b}},\ i,\ \omega,\
    \varphi_{0},\ \alpha,\ \Lambda\} \, .\label{eq:parameters}
\end{equation}
It is worth noting that the pulsar's initial position and velocity can be
derived from six initial Keplerian parameters. However, one of these parameters,
the longitude of ascending node, does not affect TOAs. Thus it is not detectable
by pulsar timing alone. So there are only five Keplerian parameters in the
timing model~\cite{BT}.

We use the generated TOAs as our observational data with Gaussian noise whose
standard variance is $\sigma_{\rm TOA}$. After mock TOAs are prepared, we fit
them with our timing model. There are residuals between the data and what are
predicted by the timing model. So, with Eq.~(\ref{eq:numbernu}) the likelihood
function of parameters in timing can be expressed as,
\begin{equation}
    P(\mathcal{N}_{i,{\rm obs}} \big| \bm{\Theta})\propto
    \exp{\left[-\frac{1}{2}\sum_{i=1}^{N_{{\rm TOA}}} \left(
    \frac{\mathcal{N}_{i,{\rm obs}}-\mathcal{N}_{i}(\bm{\Theta})}{\nu
    \sigma_{{\rm TOA}}} \right)^{2}\right]}\,,
\end{equation}
where $\nu$ is extremely well measured, $\mathcal{N}_{i,{\rm obs}}$ is the true
rotation number in which the pulse of the $i$-th TOA arrives,
$\mathcal{N}_{i}(\bm{\Theta})$ is the predicted $i$-th rotation number by the
timing model, and $N_{{\rm TOA}}$ is the total number of simulated TOAs.

With the likelihood function, we can conduct parameter estimation. We adopt the
Fisher information matrix (FIM) method and the Derivative Approximation for
LIkelihoods (DALI) method in our parameter estimation~\cite{Sellentin:2014zta,
Wang:2022kia}.  FIM is a widely used method for estimating measurement
uncertainties, and it is defined as, 
\begin{equation}
    F_{\alpha
    \beta}(\tilde{\boldsymbol{\Theta}}) \equiv \left\langle\left.\frac{\partial^2
    \mathcal{L}}{\partial \Theta^\alpha \partial
    \Theta^\beta}\right|_{\tilde{\Theta}}\right\rangle\,,
\end{equation}
where $\mathcal{L}$ is the log-likelihood function, $\tilde{\Theta}$ is the true
value of the parameter and $\langle \, \cdot \, \rangle$ means the average over the
data space. The diagonal elements of the inverse of the FIM equal to the squared
1-$\sigma$ uncertainties of the corresponding parameters.  DALI method is an
extension of FIM with higher-order terms and nonlinear dependence of parameters
in the model.  It still works well when the FIM becomes
singular~\cite{Sellentin:2014zta, Wang:2022kia}.  In addition, DALI provides a
test for the Gaussianity of posterior distributions for the FIM.  Compared with
sophisticated Markov Chain Monte Carlo (MCMC)~\cite{Hastings:1970aa} and Nested
Sampling methods~\cite{Skilling:2004,Skilling:2006gxv}, DALI still retains the
advantage of speed.  With the FIM  and the DALI methods, we can obtain the
measurement uncertainties of parameters in the timing simulation, especially for
$\alpha$ and $\Lambda$.

\section{Simulations and Results}
\label{Sec3}

In this section, we provide the specifics of simulations and the simulated
results of our work.  The main results can be divided into three parts. For each
part, we will calculate the detection ability of PSR-SMBH systems for the Yukawa
gravity in different setting.  In Section \ref{Sec3.1}, we obtain the
constraints on $\alpha$ as a function of $\Lambda$, while in Section
\ref{Sec3.2} we limit $\Lambda$ as a function of $\alpha$.  In Section
\ref{Sec3.3}, we get constraints on $\alpha$ and $\Lambda$ simultaneously.  For
the reported results, the priors are all uniform.

\subsection{Constraints on the strength of the Yukawa interaction}
\label{Sec3.1}

We consider the constraints on the strength of the Yukawa interaction $\alpha$
in the Yukawa gravity as a function of the interaction range $\Lambda$
from PSR-SMBH binary systems. The parameter set in this subsection is
$\bm{\Theta}^{(\Lambda)} \equiv \bm{\Theta} \backslash \{\Lambda\} =\{M,\ \nu,\
\dot{\nu},\ N_{0},\ e,\ P_{{b}},\ i,\ \omega,\ \varphi_{{\rm 0}},\ \alpha\}$. We
calculate the FIM by substituting source parameter values with a fiducial
$\alpha=0$. Then, the standard deviation of $\alpha$ is obtained from the
corresponding element in the inverse of the FIM, providing an estimation for the
1-$\sigma$ bound on $\alpha$.  According to Eq.~(\ref{eq:Yukawa_potential}), the
Newtonian potential is recovered when $\alpha=0$. In other words, we are
limiting $\alpha$ assuming that the Nature follows GR.  By repeating the process with
different $\Lambda$'s, we get limits on $\alpha$ as a function of $\Lambda$.

\begin{table}[tp]
\caption{Parameters of an S2-like PSR-SMBH system and its observational information.}
\centering
\renewcommand\arraystretch{1.4}
\begin{tabular}{|p{8cm}|p{4cm}|}
\hline
Parameter & Value \\
\hline
Mass of Sgr A*, $M$ ($M_{\odot}$) & $4.3\times10^{6}$\\
Spin frequency of pulsar, $\nu$ (Hz) & $2$\\
Orbital eccentricity, $e$ & $0.88$\\
Orbital period, $P_{{b}}$ (yr)  & $15.8$\\
Inclination, $i$ (deg) & $135$\\
Observational time, $T_{{\rm obs}}$ (yr) & $20$\\
Number of TOAs, $N_{{\rm TOA}}$ & $960$ \\
Timing accuracy of TOAs, $\sigma_{{\rm TOA}}$ (${\rm \mu s}$) & $100$\\
\hline
\end{tabular}
\label{Tab1_S2_par}
\end{table}

We consider a series of pulsars in our simulations. Especially, we use an
S2-like pulsar to compare with the limits from the S2
star~\cite{Gillessen:2008qv}.  The parameters of our S2-like pulsar and
observational information are given in Table\,\ref{Tab1_S2_par}.  In the view
that it is less likely to find a millisecond pulsar within the GC
region~\cite{Cordes:1996bt}, we consider young pulsars around Sgr A*, whose
rotation frequency is set to $\nu = 2\,{\rm Hz}$ in our simulations.  The
electron density in the ionized gas near the GC is high, which will seriously
influence the precision of TOA measurements, so observations at much higher
frequencies are needed to improve the precision of TOAs.  The intrinsic pulse
phase jitter can also affect the timing precision and is
frequency-dependent~\cite{Liu:2011ae,Liu:2011cka}.  Hence we need to observe
pulsars around Sgr~A* at high-frequency band, ensuring that we get pulses with
an adequate signal-to-noise ratio. Liu et al.~\cite{Liu:2011ae} predicted
the timing accuracy of a young pulsar around Sgr A*, showing that the
uncertainties of TOAs can reach below $100\,{\rm \mu s}$ with an SKA-like
telescope, assuming that the observational frequency is above 15\,GHz and the
integration time is 1 hour.  So we set $\sigma_{\rm TOA}=100\,{\rm \mu s}$ as a
realistic timing accuracy in our simulations.  As for the orbital parameters of
pulsars, Zhang et al.~\cite{Zhang:2014kva} estimated the number and the
orbital distribution of pulsars around Sgr A* by simulations. They showed that
the possibility for the existence of pulsars whose orbital semi-major axis is
less than $10^{3}\,{\rm AU}$.  So we simulate an S2-like pulsar whose Keplerian
parameters are the same as the S2 star~\cite{Gillessen:2008qv} whose orbital
semi-major axis is about $10^{3}\,{\rm AU}$.  Considering that the observational
time in pulsar timing is usually on the order of years, we set our observational
time to 20 years to ensure that it covers at least one orbital period. The
number of simulated TOAs is 960, which is about one TOA per week. 

\begin{figure}
    \centering
    \includegraphics[scale=0.76]{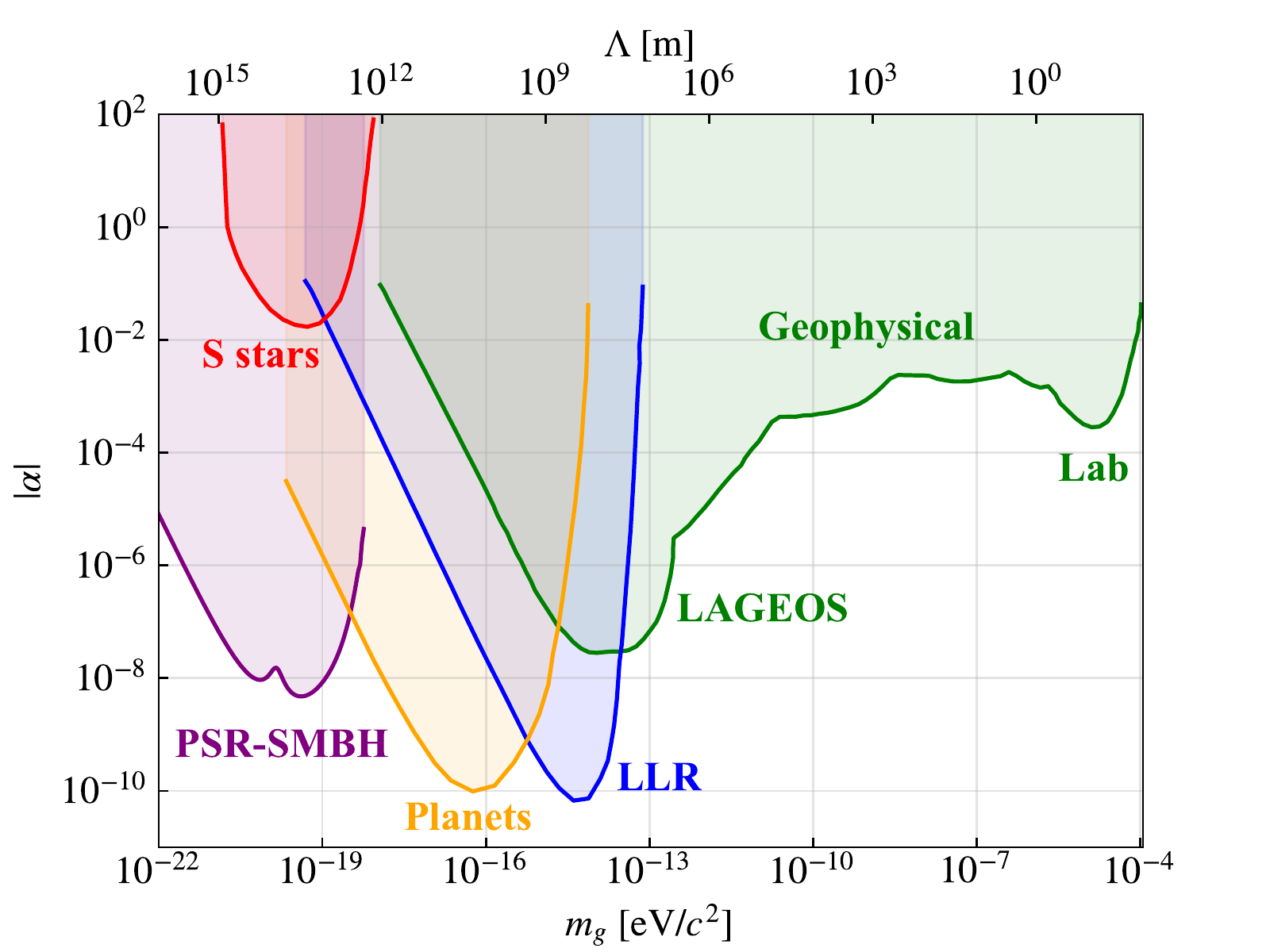}
    \caption{Constraints on the strength of the Yukawa interaction $\alpha$ at
    95\% C.L.~as a function of the range of the Yukawa interaction $\Lambda$
    from different experiments. The abscissa gives the mass of graviton via
    $m_{g}c^{2} = hc/\Lambda$. Different curves come from different experiments, and
    ``PSR-SMBH'' represents the constraints from an S2-like pulsar in
    Table~\ref{Tab1_S2_par}. The curve of ``S stars'' represents the constraints from stars
    around Sgr A*~\cite{Hees:2017aal}. The other curves come from
    Ref.~\cite{KONOPLIV2011401}.}
    \label{Fig1_limitonalpha_S2}
\end{figure}

The limits on $\alpha$ from an S2-like pulsar and limits from existing
experiments are compared in Figure~\ref{Fig1_limitonalpha_S2}. It shows that a
PSR-SMBH system has its unique advantages for tests of the Yukawa gravity at
$\Lambda$ ranging between $10^{1}$--$10^{4}\,{\rm AU}$. The parameter $\Lambda$
corresponds to the Compton wavelength of the massive graviton. It should be
compared with the other length scale in the system, in our situation the orbital
semi-major axis. The orbital semi-major axis of an S2-like pulsar is about
$10^{3}\,{\rm AU}$, which is much larger than the scale of the Solar system.  It
means that we are able to conduct tests on the Yukawa gravity when $\Lambda$ is
large and search for the fifth force in the untested parameter space. Besides,
we could test the Yukawa gravity in a strong field regime with a PSR-SMBH
system, which is impossible in the Solar system.  Compared with observations of
stars around Sgr A*, one can acquire higher ranging accuracy in pulsar timing
observations, thanks to pulsars' rotational stability, the pulsar timing
technology, and high-sensitivity telescopes.  The constraint from an S2-like
PSR-SMBH system is tighter than that of the S stars around Sgr A*.  As shown in
Figure~\ref{Fig1_limitonalpha_S2}, it will provide high-precision restrictions
on $\alpha$ when $\Lambda$ varies between $10^{1}$--$10^{4}\,{\rm AU}$. The
tightest limits on $\alpha$ at 95\% confidence level (C.L.) even reach to
$|\alpha| \lesssim 10^{-8}$.  The curve of the PSR-SMBH system has a pothole,
which is caused by the absorption of residuals by the parameter $N_{0}$ in the
timing model.  The effect of $N_{0}$ is a persistent shift of the predicted
${\mathcal{N}}_{i}\left(\bm{\Theta}\right)$ in the timing model.  If the
residuals between the observed $\mathcal{N}_{i,{\rm obs}}$ and the predicted
$\mathcal{N}_{i}(\bm{\Theta})$ are relatively flat or monotone, the absorption
by $N_{0}$ will be obvious.  In our observation, the observational time span is
set to $20\,{\rm yr}$ and the orbital period of the S2-like pulsar is
$15.8\,{\rm yr}$, which means that we only observe a little more than one
period.  If there is a set of parameters that makes the residuals relatively
flat or monotone, the accuracy will decrease substantially at that point.

\begin{figure}
    \centering
    \includegraphics[scale=0.48]{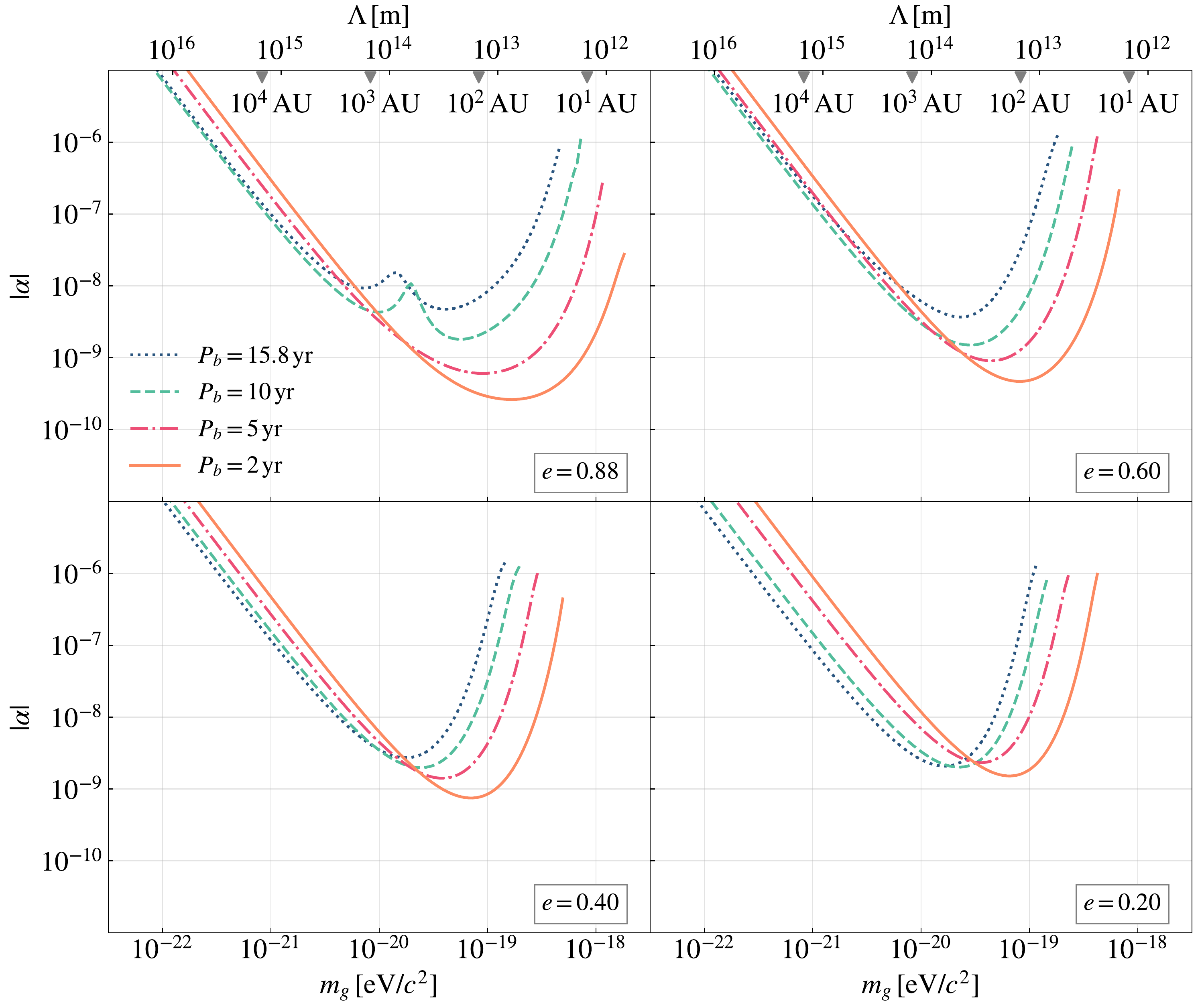}
    \caption{Constraints on the strength of the Yukawa interaction $\alpha$ at
    68\% C.L.~from pulsars with different orbital periods and eccentricities.
    Different colored curves correspond to different orbital periods and
    different panels correspond to different eccentricities that are shown in
    the lower right corner. Other orbital parameters and observational
    information are the same as that in Table~\ref{Tab1_S2_par}.}
    \label{Fig2_limitonalpha_eandPb}
\end{figure}

Besides the restrictions on $\alpha$ from an S2-like pulsar around Sgr A*, we
also perform simulations with different orbital parameters.  Figure
\ref{Fig2_limitonalpha_eandPb} shows how the orbital period and eccentricity of
pulsars influence the constraints on $\alpha$.  Different colored curves
correspond to different orbital periods, and different panels correspond to
different eccentricities.  For a selected system, the constraint on $\alpha$
usually becomes tightest when $\Lambda$ is close to the length scale of the
system because, in this case, the Yukawa gravity has a significant deviation
from the inverse-square law.  A shorter orbital period corresponds to a smaller
orbital semi-major axis, so the tightest bound on $\alpha$ will be obtained
where $\Lambda$ is smaller.  The orbtial eccentricity $e$ affects the range of
distance covered by the system.  A larger $e$ means that the distance between
the pulsar and the SMBH varies more, so we can have decent bounds on $\alpha$ at
a larger range of $\Lambda$. This corresponds to a wider pocket of the curves in
Figure~\ref{Fig2_limitonalpha_eandPb}.  In the figure, there are also potholes
on curves from pulsars with high eccentricities and long orbital periods, caused
by the absorption of residuals by the parameter $N_{0}$.  As discussed earlier,
when compared with short orbital period PSR-SMBH systems, within $T_{\rm obs} =
20$\,yr we can only observe one or two orbital periods for long orbital period
PSR-SMBH systems, which means that the residuals from long orbital period
PSR-SMBH systems may lack periodicity and become monotonous, and thus can be
absorbed by the parameter $N_{0}$.  Combined with the limited observational
time, the absorption of the residuals by $N_{0}$ is more obvious for pulsars
with high eccentricities. For pulsars with high eccentricities, the residuals
caused by the Yukawa gravity will be mainly distributed at the part of
periastron, showing a peak; for pulsars with low eccentricities, the residuals
will be distributed across the orbital phases, showing a smooth curve. If the
periastron of one orbital period is missed, for pulsars with high
eccentricities, the residuals caused by the Yukawa gravity will be almost flat
in that orbital period. But the absence of observation for the periastron has
relatively little effects on pulsars with low eccentricities. So the absorption
of the residuals by $N_{0}$ is more obvious for pulsars with high
eccentricities. 

\subsection{Constraints on the range of the Yukawa interaction}
\label{Sec3.2}

In this subsection we limit the range of the Yukawa interaction $\Lambda$, which
is related to the graviton mass $m_g$.  The range of the Yukawa interaction
$\Lambda$ corresponds to the Compton wavelength of the graviton, so the graviton
mass can be expressed as~\cite{Zakharov:2016lzv,Zakharov:2018cbj},
\begin{equation}
    m_{g}c^{2}=\frac{hc}{\Lambda} \,,
\end{equation}
where $h$ is the Planck constant.  In GR, $m_g$ equals to zero and $\Lambda \to
\infty$. The massive graviton will contribute an additional periastron advance
rate, which is also the main source of timing residuals. Through the method of
orbital perturbation~\cite{Gravity}, when the orbital semi-major axis $a_{\rm
orb} \ll \Lambda$, the periastron advance from the Yukawa potential for each
orbit can be expressed as~\cite{Zakharov:2018cbj},
\begin{equation}
    \Delta \omega =
    \frac{\pi\alpha\sqrt{1-e^{2}}}{1+\alpha}\frac{a_{\rm orb}^2}{\Lambda^2}
    \label{eq_advance_of_periastron} \,.
\end{equation}
According to Eq.~(\ref{eq_advance_of_periastron}), when $\alpha$ changes from
$\alpha=1$ to infinity,  $\Delta \omega$ will only double. So with a given
observational constraint on $\Delta \omega$, the limit on $1/\Lambda^2 \sim
m_g^2$ will change by the same amount.  Therefore, here we fix $\alpha=1$ for an
example. The limits on $m_{g}$ with another $\alpha$ can be easily estimated
with Eq.~(\ref{eq_advance_of_periastron}). Now the set of parameters is
$\bm{\Theta}^{(\alpha)} \equiv \bm{\Theta} \backslash \{ \alpha \} =\{M,\ \nu,\
\dot{\nu},\ N_{0},\ e,\ P_{{b}},\ i,\ \omega,\ \varphi_{{\rm 0}},\ \Lambda\}$.
GR can be recovered when the length scale $\Lambda$ tends to infinity, so we
choose the fiducial value of $1/\Lambda$ to be zero, which means that we are
limiting $m_{g}$ with a fiducial $m_{g}=0$.  In the process of calculating the
FIM, it is needed to calculate the first derivatives of $\mathcal{N}_{i}$ with
respect to the model parameters. When $r\ll \Lambda$, the main effects on
$\mathcal{N}_{i}$ by the Yukawa gravity are from the extra periastron
advance~(\ref{eq_advance_of_periastron}) from the Yukawa potential. The extra
periastron advance is proportional to $1/\Lambda^2$. If we take the derivative
of $1/\Lambda$ and set $1/\Lambda$ to 0, this term will equal to zero and thus
the first derivative of $\mathcal{N}_{i}$ with respect to $1/\Lambda$ will also
equal to zero, which causes a situation where the FIM is approximately singular
when the fiducial value for $1/\Lambda$ is zero. In this case, the FIM can not
give the valid parameter estimation. So we adopt the DALI method to get the
standard deviation of $m_{g}$. We use \texttt{emcee}~\cite{2013PASP..125..306F}
and \texttt{corner}~\cite{corner} packages in our DALI calculation. 

\begin{figure}
    \centering
    \includegraphics[scale=0.76]{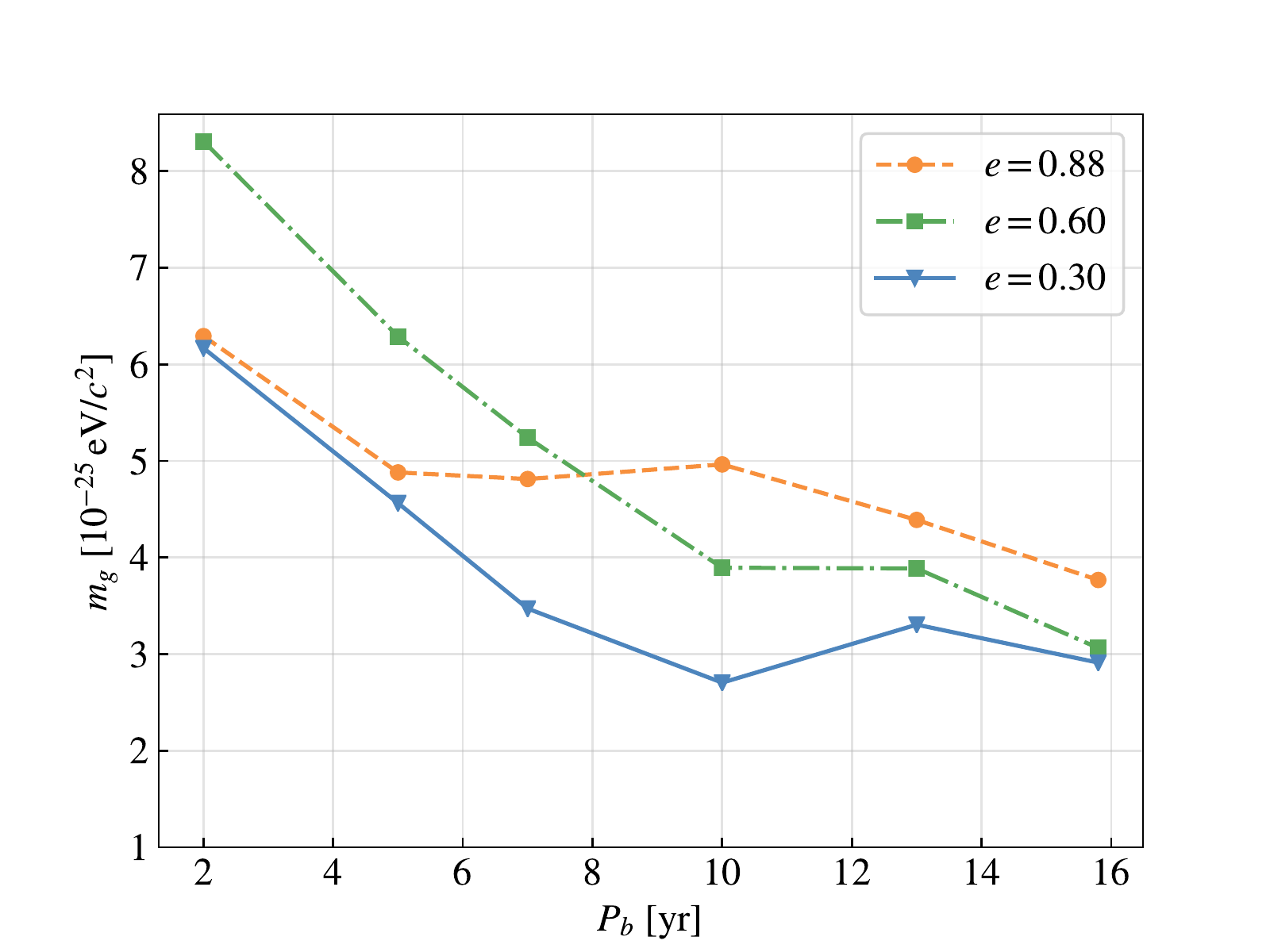}
    \caption{Constraints on the graviton mass at 68\% C.L.\ with PSR-SMBH
    systems as a function of orbital period. Different color means different
    orbital eccentricities of pulsar orbits. Other orbital parameters and
    observational information are the same as that in Table \ref{Tab1_S2_par}.}
    \label{Fig3_gravitonmass}
\end{figure}

We consider limits on $m_{g}$ from a number of pulsars with different orbital
periods and eccentricities. Other parameters of these pulsars and the
observational information are the same as that in Table~\ref{Tab1_S2_par}.  The
constraints on the graviton mass $m_{g}$ from different PSR-SMBH systems are
shown in Figure~\ref{Fig3_gravitonmass}.  As we can see, the constraints on
$m_{g}$ can be as small as $10^{-25}\,{\rm eV}/c^{2}$ at 68\%~C.L. from an
S2-like pulsar.  Compared with the bounds on $m_{g}$ from different experiments
in Section \ref{Sec1}, the limit from PSR-SMBH systems is more competitive,
thanks to the high accuracy of pulsar timing technique. Besides,  the PSR-SMBH
systems are providing limits on the graviton mass in a strong-field regime. With
the orbital semi-major axis, $a_{\rm orb}\propto P_{{b}}^{2/3} M^{1/3}$, we get
the orbital period average for the periastron advance rate from the Yukawa
gravity as $\dot{\omega} \equiv \Delta \omega / P_{{b}} \propto
P_{{b}}^{1/3}M^{2/3}$ according to Eq.~(\ref{eq_advance_of_periastron}).  Due to
the huge mass of Sgr~A* and the orbital periods of pulsars $P_b \gtrsim 1\,$yr
under consideration, the rate of periastron advance will be significant.
Overall, the PSR-SMBH systems provide an unrivalled opportunity to bound
graviton mass. 

In Figure~\ref{Fig3_gravitonmass}, for pulsars with a same orbital eccentricity,
a larger $P_{b}$  provides a tighter limit on $m_{g}$, which is a major trend
according to the analysis  of the periastron advance rate.  But the curves do
not show strict monotonicity because multiple factors affect the bounds on
$m_g$.  First, in the case of pulsar timing, given an observational span $T_{\rm
obs}$, a larger $P_{b}$ means worse measurement accuracy of orbital parameters,
which worsens the limit on $m_{g}$.  Second, the absorption of residuals by the
parameter $N_{0}$ will be significant for pulsars with long orbital periods and
large eccentricities, which is detrimental to the bound.  Finally, from the
viewpoint of the actual observation, a larger $P_{b}$ is not always better for
the constraint. For a pulsar with an extremely long $P_{b}$, it is difficult to
cover  one orbital period within the limited observational time, which leads to
a looser bound on $m_{g}$. External perturbations will also be relevant, which
are hard to model. For pulsars with both long orbital periods and high
eccentricities, if we are unfortunate to miss the part of the periastron, the
absence of that will greatly affect the bound.  For pulsars with a same orbital
period, the $m_g$ bound also depends on the eccentricity. According to
Eq.~(\ref{eq_advance_of_periastron}), for a same $P_{b}$, the smaller the
eccentricity is, the greater the advance of periastron is.  At the same time, a
small orbital eccentricity will reduce the measurement precision of the advance
of periastron in pulsar timing. Combining these effects in our simulation, we
finally get the results in Figure~\ref{Fig3_gravitonmass}, showing the
dependence of the $m_g$ bound on the orbital period and orbital eccentricity.

\begin{figure}
    \centering
    \includegraphics[scale=0.56]{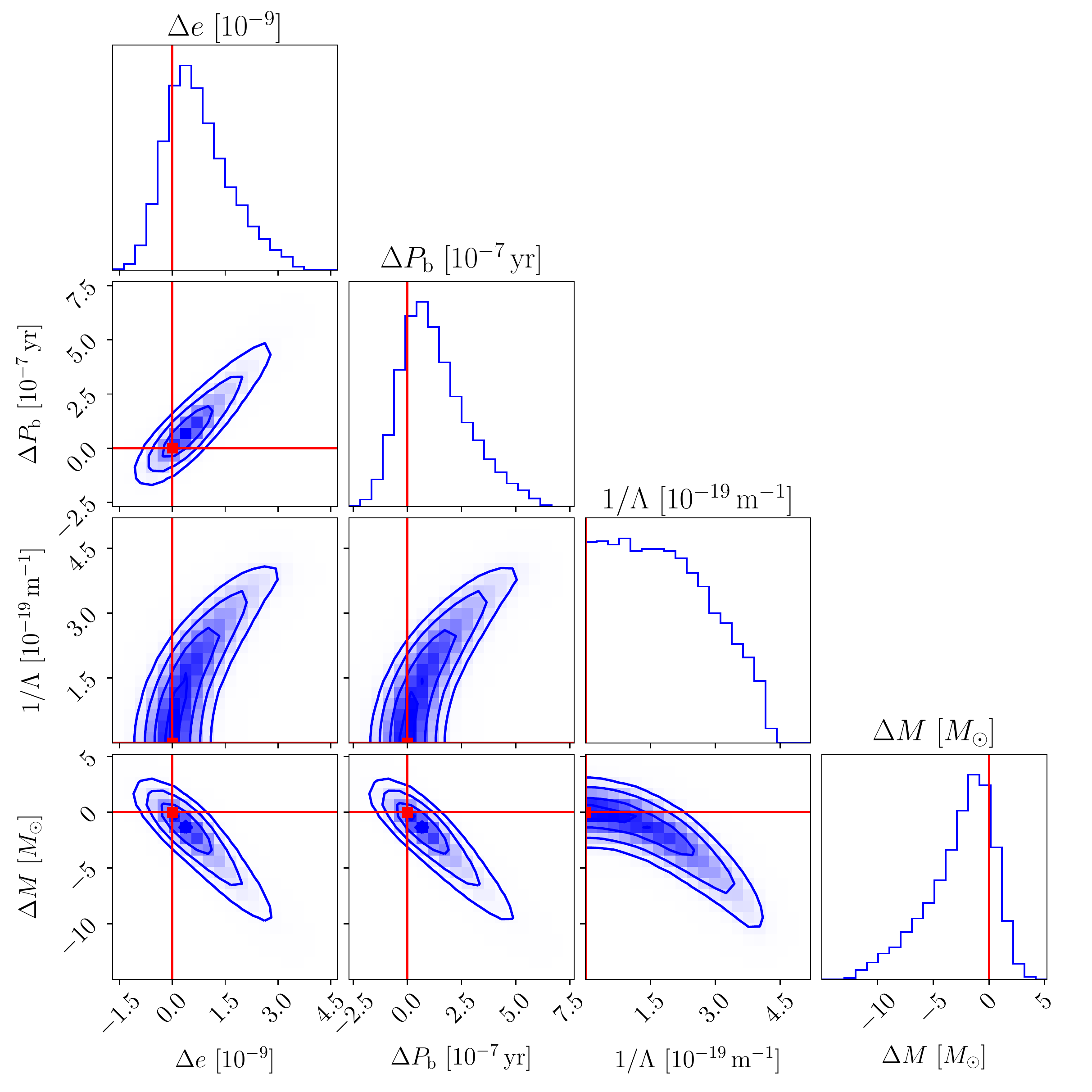}
    \caption{The posterior distribution of some parameters from the DALI method
    for an S2-like pulsar. The  contours are drawn at the $1$-$\sigma$,
    $2$-$\sigma$ and $3$-$\sigma$ confidence levels. $\Delta e$ represents the
    difference between  $e$ and its truth value in Table \ref{Tab1_S2_par}.
    $\Delta P_{{b}}$ and $\Delta M$ are given in the same way.  Truth values of
    $\Delta e$, $\Delta P_{{b}}$, $\Delta M$, and $1/\Lambda$ are all zero and
    indicated as red lines. }
    \label{Fig4_gravitonmass_DALI}
\end{figure}

In Figure \ref{Fig4_gravitonmass_DALI} we show the posterior distribution of
some parameters of an S2-like pulsar. The distortion in the ellipses in the
posterior contours reflects the effects of higher order terms and the prior that
$\Lambda>0$. Since the FIM can not give the valid posterior distribution, we
have used the DALI method. 

\subsection{Simultaneous bounds on $\alpha$ and $\Lambda$}
\label{Sec3.3}

\begin{figure}
    \centering
    \includegraphics[scale=0.6]{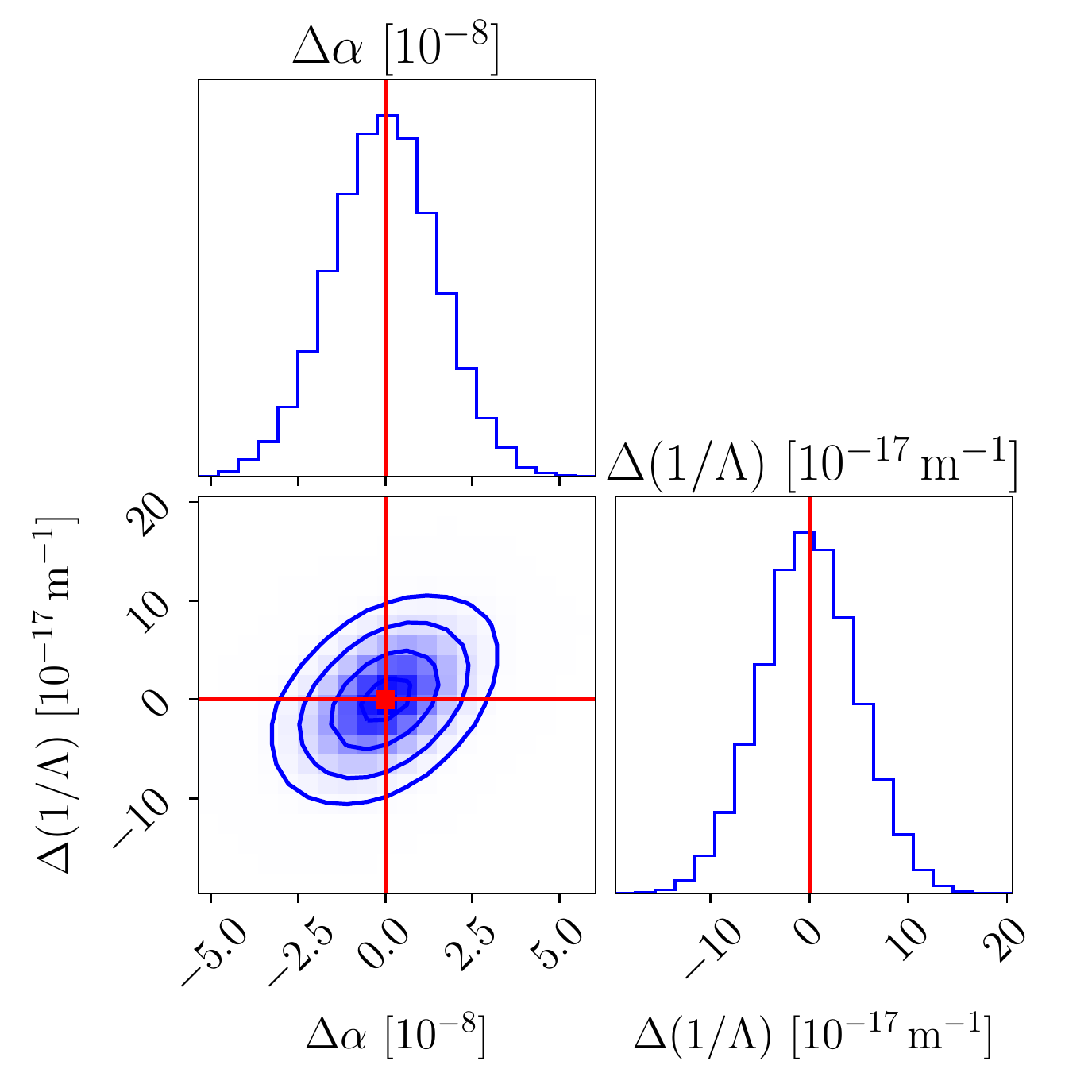}
    \caption{Distribution of parameter posteriors in the Yukawa gravity. The
    contours are drawn at the $1$-$\sigma$, $2$-$\sigma$ and $3$-$\sigma$
    confidence levels. The distribution of posteriors is obtained by the DALI
    method. $\Delta \alpha$ and $\Delta (1/\Lambda)$ are the difference between
    parameters and their truth values, shown in Eq.~(\ref{Sec3.1_Yukawa_par}).
    So the truth values of $\Delta \alpha$ and $\Delta (1/\Lambda)$ are zero and
    indicated as red lines.}
    \label{Fig5_broken_DALI}
\end{figure}

In the above subsections, we have limited one of the Yukawa gravity parameters
with the other one fixed.  But if the Nature indeed follows the Yukawa gravity,
i.e.~$\alpha$ is non-zero and $\Lambda$ is finite, we need to limit both
parameters simultaneously.  However, there exist some difficulties to do so.  If
the length scale of system is much less than $\Lambda$, according to
Eq.~(\ref{eq_advance_of_periastron}), the two parameters $\alpha$ and $\Lambda$
are highly degenerate.  But if the length scale of the tested system is similar
to $\Lambda$, we can detect higher order effects of $1/\Lambda$ and obtain
decent limits on $\alpha$ and $\Lambda$ at the same time.  So for a PSR-SMBH
system with its length scale comparable to $\Lambda$, it is expected to break
the degeneracy of $\alpha$ and $\Lambda$. 

In this subsection, we test the ability of the PSR-SMBH system to measure
$\alpha$ and $\Lambda$ simultaneously when the gravity is assumed to be the
Yukawa gravity. We will calculate the standard deviations of $\alpha$ and
$\Lambda$ simultaneously at fiducial values that $\alpha$ and $1/\Lambda$ are
nonzero.  In other words, we generate TOAs assuming that the gravity is the Yukawa
gravity. Then we try to test whether we can measure $\alpha$ and $\Lambda$
simultaneously through a PSR-SMBH system.  The set of parameters is
$\bm{\Theta}=\{M,\ \nu,\ \dot{\nu},\ N_{0},\ e,\ P_{{b}},\ i,\ \omega,\
\varphi_{{\rm 0}},\ \alpha,\ \Lambda\}$.  We adopt the DALI method to get more
accurate parameter posteriors.

We choose an S2-like pulsar as an example, and the observational information is
given in Table~\ref{Tab1_S2_par}. We take one set of fiducial values of $\alpha$
and $\Lambda$ for illustration. The selection of fiducial values is based on the
following considerations. First, we should be able to detect the deviation from
the fiducial values, which means that we should choose the point above the curve
of PSR-SMBH in Figure~\ref{Fig1_limitonalpha_S2}.  Second, we would prefer that
$\Lambda$ is similar to the orbital semi-major axis of an S2-like pulsar.  We
are more likely to break the degeneracy of $\alpha$ and $\Lambda$ in this case.
Finally, we choose the fiducial values as,
\begin{subequations}\label{eq3.3:}
\begin{align}
\alpha & = 10^{-6} \, ,
\\
1/\Lambda & = 10^{-14}\,{\rm m}^{-1} \, .
\end{align}
\label{Sec3.1_Yukawa_par}
\end{subequations}

Through parameter estimation on the simulated pulsar timing data, the
distribution of two parameters of the Yukawa gravity is shown in Figure
\ref{Fig5_broken_DALI} using the DALI method. The posterior contours are almost
elliptical with parameters in Eq.~(\ref{eq3.3:}). We can prove that the
posterior distributions are approximately Gaussian.  From the figure, an S2-like
PSR-SMBH system can give the following 1-$\sigma$ measurement uncertainties on
the Yukawa gravity parameters,
\begin{subequations}
\begin{align}
\sigma_{\alpha} & = 1.5\times10^{-8} \, ,
\\
\sigma_{1/\Lambda} & = 4.8\times10^{-17}\,{\rm m}^{-1} \, .
\end{align}
\end{subequations}
As we can see, we have $\sigma_{\alpha}/\alpha=1.5\%$ and
$\sigma_{1/\Lambda}/(1/\Lambda)=0.48\%$, which means that we can make a
simultaneous measurement of $\alpha$ and $\Lambda$ in this artificially designed
example. It demonstrates the prospects of PSR-SMBH systems measuring the Yukawa
gravity parameters when the theory parameters happen to be optimal.

\section{Conclusion and Outlook}
\label{Sec4}

In this paper, we discuss the test of the Yukawa gravity by timing a pulsar
around Sgr A*.  Considering an S2-like pulsar with 960 TOAs in a $20$-${\rm yr}$
observation, with  an assumed realistic timing accuracy  $\sigma_{\rm TOA} =
100\,{\rm \mu s}$, we can expect to achieve the following.
\begin{enumerate}[(i)]
	\item When the range of the Yukawa interaction $\Lambda$ varies between
	$10^{1}$--$10^{4}\,{\rm AU}$, the constraints on the strength of the Yukawa
	interaction $|\alpha|$ will range between $10^{-8}$ and $10^{-6}$ (see
	Figure~\ref{Fig1_limitonalpha_S2}). Compared with other existing
	experiments, pulsars around Sgr A* will significantly shrink the parameter
	space of the Yukawa gravity in the region where $\Lambda =
	10^{1}$--$10^{4}\,{\rm AU}$.
	\item By converting the range parameter into the graviton mass $m_{g}$, the
	upper limit on  $m_{g}$ can reach as small as $10^{-25}\,{\rm eV}/c^2$, which is
	tighter than the results from the Solar system and observation of stars
	around Sgr A*.
	\item If the gravity indeed follows the Yukawa gravity and $\Lambda$ is
	similar to the length scale of a PSR-SMBH system, it has the potential to
	break the degeneracy of the Yukawa gravity parameters, namely the strength
	parameter $\alpha$ and the range parameter $\Lambda$ simultaneously.
\end{enumerate}

According to astrophysical estimates, it is likely to find a pulsar around Sgr
A* by the SKA~\cite{Cordes:1996bt,Smits:2008cf,Wharton:2011dv,Bower:2018mta}.
Moreover, the assumed timing accuracy in this work, $\sigma_{\rm TOA} =
100\,{\rm \mu s}$, is realistic for next-generation SKA-like
telescopes~\cite{Liu:2011ae}, and the assumption that TOAs' detection cadence is
once per week is relatively conservative. In the future, it is promising to
achieve the estimated accuracy in SKA-like projects if suitable pulsars around
Sgr A* can be detected.  Besides, a deeper understanding of the environment
around Sgr A* will provide a more suitable PSR-SMBH timing model.  We expect
that we can detect useful pulsars around Sgr A*  in the future, and promote
gravity tests in the strong field in a profound way. 

\acknowledgments

This work was supported by the National SKA Program of China (2020SKA0120300),
the National Natural Science Foundation of China (11991053, 12203072, 11975027,
11721303), the Max Planck Partner Group Program funded by the Max Planck
Society, and the High-Performance Computing Platform of Peking University.  
Y.D., Z.H., and Z.W.\ are respectively supported by the National Training Progam
of Innovation for Undergraduates, the Principal's Fund for the Undergraduate
Student Research Study, and the Hui-Chun Chin and Tsung-Dao Lee Chinese
Undergraduate Research Endowment (Chun-Tsung Endowment) at Peking University.
X.M.\ is supported by the FAST Postdoctoral Fellowship at the National
Astronomical Observatories, Chinese Academy of Sciences.










\bibliographystyle{apsrev4-1}
\bibliography{refs}

\begin{thebibliography}{84}%
\makeatletter
\providecommand \@ifxundefined [1]{%
 \@ifx{#1\undefined}
}%
\providecommand \@ifnum [1]{%
 \ifnum #1\expandafter \@firstoftwo
 \else \expandafter \@secondoftwo
 \fi
}%
\providecommand \@ifx [1]{%
 \ifx #1\expandafter \@firstoftwo
 \else \expandafter \@secondoftwo
 \fi
}%
\providecommand \natexlab [1]{#1}%
\providecommand \enquote  [1]{``#1''}%
\providecommand \bibnamefont  [1]{#1}%
\providecommand \bibfnamefont [1]{#1}%
\providecommand \citenamefont [1]{#1}%
\providecommand \href@noop [0]{\@secondoftwo}%
\providecommand \href [0]{\begingroup \@sanitize@url \@href}%
\providecommand \@href[1]{\@@startlink{#1}\@@href}%
\providecommand \@@href[1]{\endgroup#1\@@endlink}%
\providecommand \@sanitize@url [0]{\catcode `\\12\catcode `\$12\catcode
  `\&12\catcode `\#12\catcode `\^12\catcode `\_12\catcode `\%12\relax}%
\providecommand \@@startlink[1]{}%
\providecommand \@@endlink[0]{}%
\providecommand \url  [0]{\begingroup\@sanitize@url \@url }%
\providecommand \@url [1]{\endgroup\@href {#1}{\urlprefix }}%
\providecommand \urlprefix  [0]{URL }%
\providecommand \Eprint [0]{\href }%
\providecommand \doibase [0]{http://dx.doi.org/}%
\providecommand \selectlanguage [0]{\@gobble}%
\providecommand \bibinfo  [0]{\@secondoftwo}%
\providecommand \bibfield  [0]{\@secondoftwo}%
\providecommand \translation [1]{[#1]}%
\providecommand \BibitemOpen [0]{}%
\providecommand \bibitemStop [0]{}%
\providecommand \bibitemNoStop [0]{.\EOS\space}%
\providecommand \EOS [0]{\spacefactor3000\relax}%
\providecommand \BibitemShut  [1]{\csname bibitem#1\endcsname}%
\let\auto@bib@innerbib\@empty
\bibitem [{\citenamefont {Will}(2014)}]{Will:2014kxa}%
  \BibitemOpen
  \bibfield  {author} {\bibinfo {author} {\bibfnamefont {C.~M.}\ \bibnamefont
  {Will}},\ }\href {\doibase 10.12942/lrr-2014-4} {\bibfield  {journal}
  {\bibinfo  {journal} {Living Rev. Rel.}\ }\textbf {\bibinfo {volume} {17}},\
  \bibinfo {pages} {4} (\bibinfo {year} {2014})},\ \Eprint
  {http://arxiv.org/abs/1403.7377} {arXiv:1403.7377 [gr-qc]} \BibitemShut
  {NoStop}%
\bibitem [{\citenamefont {Berti}\ \emph {et~al.}(2015)\citenamefont {Berti}
  \emph {et~al.}}]{Berti:2015itd}%
  \BibitemOpen
  \bibfield  {author} {\bibinfo {author} {\bibfnamefont {E.}~\bibnamefont
  {Berti}} \emph {et~al.},\ }\href {\doibase 10.1088/0264-9381/32/24/243001}
  {\bibfield  {journal} {\bibinfo  {journal} {Class. Quant. Grav.}\ }\textbf
  {\bibinfo {volume} {32}},\ \bibinfo {pages} {243001} (\bibinfo {year}
  {2015})},\ \Eprint {http://arxiv.org/abs/1501.07274} {arXiv:1501.07274
  [gr-qc]} \BibitemShut {NoStop}%
\bibitem [{\citenamefont {Bertone}\ and\ \citenamefont
  {Hooper}(2018)}]{Bertone:2016nfn}%
  \BibitemOpen
  \bibfield  {author} {\bibinfo {author} {\bibfnamefont {G.}~\bibnamefont
  {Bertone}}\ and\ \bibinfo {author} {\bibfnamefont {D.}~\bibnamefont
  {Hooper}},\ }\href {\doibase 10.1103/RevModPhys.90.045002} {\bibfield
  {journal} {\bibinfo  {journal} {Rev. Mod. Phys.}\ }\textbf {\bibinfo {volume}
  {90}},\ \bibinfo {pages} {045002} (\bibinfo {year} {2018})},\ \Eprint
  {http://arxiv.org/abs/1605.04909} {arXiv:1605.04909 [astro-ph.CO]}
  \BibitemShut {NoStop}%
\bibitem [{\citenamefont {Debono}\ and\ \citenamefont
  {Smoot}(2016)}]{Debono:2016vkp}%
  \BibitemOpen
  \bibfield  {author} {\bibinfo {author} {\bibfnamefont {I.}~\bibnamefont
  {Debono}}\ and\ \bibinfo {author} {\bibfnamefont {G.~F.}\ \bibnamefont
  {Smoot}},\ }\href {\doibase 10.3390/universe2040023} {\bibfield  {journal}
  {\bibinfo  {journal} {Universe}\ }\textbf {\bibinfo {volume} {2}},\ \bibinfo
  {pages} {23} (\bibinfo {year} {2016})},\ \Eprint
  {http://arxiv.org/abs/1609.09781} {arXiv:1609.09781 [gr-qc]} \BibitemShut
  {NoStop}%
\bibitem [{\citenamefont {Moffat}(2004)}]{Moffat:2004nw}%
  \BibitemOpen
  \bibfield  {author} {\bibinfo {author} {\bibfnamefont {J.~W.}\ \bibnamefont
  {Moffat}},\ }\href@noop {} {\  (\bibinfo {year} {2004})},\ \Eprint
  {http://arxiv.org/abs/astro-ph/0403266} {arXiv:astro-ph/0403266} \BibitemShut
  {NoStop}%
\bibitem [{\citenamefont {Saridakis}\ \emph {et~al.}(2021)\citenamefont
  {Saridakis}, \citenamefont {Lazkoz}, \citenamefont {Salzano}, \citenamefont
  {Vargas~Moniz}, \citenamefont {Capozziello}, \citenamefont
  {Beltr\'an~Jim\'enez}, \citenamefont {De~Laurentis},\ and\ \citenamefont
  {Olmo}}]{Saridakis:2021vue}%
  \BibitemOpen
  \bibinfo {editor} {\bibfnamefont {E.~N.}\ \bibnamefont {Saridakis}}, \bibinfo
  {editor} {\bibfnamefont {R.}~\bibnamefont {Lazkoz}}, \bibinfo {editor}
  {\bibfnamefont {V.}~\bibnamefont {Salzano}}, \bibinfo {editor} {\bibfnamefont
  {P.}~\bibnamefont {Vargas~Moniz}}, \bibinfo {editor} {\bibfnamefont
  {S.}~\bibnamefont {Capozziello}}, \bibinfo {editor} {\bibfnamefont
  {J.}~\bibnamefont {Beltr\'an~Jim\'enez}}, \bibinfo {editor} {\bibfnamefont
  {M.}~\bibnamefont {De~Laurentis}}, \ and\ \bibinfo {editor} {\bibfnamefont
  {G.~J.}\ \bibnamefont {Olmo}},\ eds.,\ \href {\doibase
  10.1007/978-3-030-83715-0} {\emph {\bibinfo {title} {{Modified Gravity and
  Cosmology}}}}\ (\bibinfo  {publisher} {Springer},\ \bibinfo {year}
  {2021})\BibitemShut {NoStop}%
\bibitem [{\citenamefont {Hinterbichler}(2012)}]{Hinterbichler:2011tt}%
  \BibitemOpen
  \bibfield  {author} {\bibinfo {author} {\bibfnamefont {K.}~\bibnamefont
  {Hinterbichler}},\ }\href {\doibase 10.1103/RevModPhys.84.671} {\bibfield
  {journal} {\bibinfo  {journal} {Rev. Mod. Phys.}\ }\textbf {\bibinfo {volume}
  {84}},\ \bibinfo {pages} {671} (\bibinfo {year} {2012})},\ \Eprint
  {http://arxiv.org/abs/1105.3735} {arXiv:1105.3735 [hep-th]} \BibitemShut
  {NoStop}%
\bibitem [{\citenamefont {van Dam}\ and\ \citenamefont
  {Veltman}(1970)}]{vanDam:1970vg}%
  \BibitemOpen
  \bibfield  {author} {\bibinfo {author} {\bibfnamefont {H.}~\bibnamefont {van
  Dam}}\ and\ \bibinfo {author} {\bibfnamefont {M.~J.~G.}\ \bibnamefont
  {Veltman}},\ }\href {\doibase 10.1016/0550-3213(70)90416-5} {\bibfield
  {journal} {\bibinfo  {journal} {Nucl. Phys. B}\ }\textbf {\bibinfo {volume}
  {22}},\ \bibinfo {pages} {397} (\bibinfo {year} {1970})}\BibitemShut
  {NoStop}%
\bibitem [{\citenamefont {Zakharov}(1970)}]{Zakharov:1970cc}%
  \BibitemOpen
  \bibfield  {author} {\bibinfo {author} {\bibfnamefont {V.~I.}\ \bibnamefont
  {Zakharov}},\ }\href@noop {} {\bibfield  {journal} {\bibinfo  {journal} {JETP
  Lett.}\ }\textbf {\bibinfo {volume} {12}},\ \bibinfo {pages} {312} (\bibinfo
  {year} {1970})}\BibitemShut {NoStop}%
\bibitem [{\citenamefont {Vainshtein}(1972)}]{Vainshtein:1972sx}%
  \BibitemOpen
  \bibfield  {author} {\bibinfo {author} {\bibfnamefont {A.~I.}\ \bibnamefont
  {Vainshtein}},\ }\href {\doibase 10.1016/0370-2693(72)90147-5} {\bibfield
  {journal} {\bibinfo  {journal} {Phys. Lett. B}\ }\textbf {\bibinfo {volume}
  {39}},\ \bibinfo {pages} {393} (\bibinfo {year} {1972})}\BibitemShut
  {NoStop}%
\bibitem [{\citenamefont {de~Rham}(2014)}]{deRham:2014zqa}%
  \BibitemOpen
  \bibfield  {author} {\bibinfo {author} {\bibfnamefont {C.}~\bibnamefont
  {de~Rham}},\ }\href {\doibase 10.12942/lrr-2014-7} {\bibfield  {journal}
  {\bibinfo  {journal} {Living Rev. Rel.}\ }\textbf {\bibinfo {volume} {17}},\
  \bibinfo {pages} {7} (\bibinfo {year} {2014})},\ \Eprint
  {http://arxiv.org/abs/1401.4173} {arXiv:1401.4173 [hep-th]} \BibitemShut
  {NoStop}%
\bibitem [{\citenamefont {Borka}\ \emph {et~al.}(2013)\citenamefont {Borka},
  \citenamefont {Jovanovi\'c}, \citenamefont {Jovanovi\'c},\ and\ \citenamefont
  {Zakharov}}]{Borka:2013dba}%
  \BibitemOpen
  \bibfield  {author} {\bibinfo {author} {\bibfnamefont {D.}~\bibnamefont
  {Borka}}, \bibinfo {author} {\bibfnamefont {P.}~\bibnamefont {Jovanovi\'c}},
  \bibinfo {author} {\bibfnamefont {V.~B.}\ \bibnamefont {Jovanovi\'c}}, \ and\
  \bibinfo {author} {\bibfnamefont {A.~F.}\ \bibnamefont {Zakharov}},\ }\href
  {\doibase 10.1088/1475-7516/2013/11/050} {\bibfield  {journal} {\bibinfo
  {journal} {JCAP}\ }\textbf {\bibinfo {volume} {11}},\ \bibinfo {pages} {050}
  (\bibinfo {year} {2013})},\ \Eprint {http://arxiv.org/abs/1311.1404}
  {arXiv:1311.1404 [astro-ph.GA]} \BibitemShut {NoStop}%
\bibitem [{\citenamefont {Zakharov}\ \emph {et~al.}(2016)\citenamefont
  {Zakharov}, \citenamefont {Jovanovic}, \citenamefont {Borka},\ and\
  \citenamefont {Jovanovic}}]{Zakharov:2016lzv}%
  \BibitemOpen
  \bibfield  {author} {\bibinfo {author} {\bibfnamefont {A.~F.}\ \bibnamefont
  {Zakharov}}, \bibinfo {author} {\bibfnamefont {P.}~\bibnamefont {Jovanovic}},
  \bibinfo {author} {\bibfnamefont {D.}~\bibnamefont {Borka}}, \ and\ \bibinfo
  {author} {\bibfnamefont {V.~B.}\ \bibnamefont {Jovanovic}},\ }\href {\doibase
  10.1088/1475-7516/2016/05/045} {\bibfield  {journal} {\bibinfo  {journal}
  {JCAP}\ }\textbf {\bibinfo {volume} {05}},\ \bibinfo {pages} {045} (\bibinfo
  {year} {2016})},\ \Eprint {http://arxiv.org/abs/1605.00913} {arXiv:1605.00913
  [gr-qc]} \BibitemShut {NoStop}%
\bibitem [{\citenamefont {Zakharov}\ \emph {et~al.}(2018)\citenamefont
  {Zakharov}, \citenamefont {Jovanovi\'c}, \citenamefont {Borka},\ and\
  \citenamefont {Borka~Jovanovi\'c}}]{Zakharov:2018cbj}%
  \BibitemOpen
  \bibfield  {author} {\bibinfo {author} {\bibfnamefont {A.~F.}\ \bibnamefont
  {Zakharov}}, \bibinfo {author} {\bibfnamefont {P.}~\bibnamefont
  {Jovanovi\'c}}, \bibinfo {author} {\bibfnamefont {D.}~\bibnamefont {Borka}},
  \ and\ \bibinfo {author} {\bibfnamefont {V.}~\bibnamefont
  {Borka~Jovanovi\'c}},\ }\href {\doibase 10.1088/1475-7516/2018/04/050}
  {\bibfield  {journal} {\bibinfo  {journal} {JCAP}\ }\textbf {\bibinfo
  {volume} {04}},\ \bibinfo {pages} {050} (\bibinfo {year} {2018})},\ \Eprint
  {http://arxiv.org/abs/1801.04679} {arXiv:1801.04679 [gr-qc]} \BibitemShut
  {NoStop}%
\bibitem [{\citenamefont {Talmadge}\ \emph {et~al.}(1988)\citenamefont
  {Talmadge}, \citenamefont {Berthias}, \citenamefont {Hellings},\ and\
  \citenamefont {Standish}}]{Talmadge:1988qz}%
  \BibitemOpen
  \bibfield  {author} {\bibinfo {author} {\bibfnamefont {C.}~\bibnamefont
  {Talmadge}}, \bibinfo {author} {\bibfnamefont {J.~P.}\ \bibnamefont
  {Berthias}}, \bibinfo {author} {\bibfnamefont {R.~W.}\ \bibnamefont
  {Hellings}}, \ and\ \bibinfo {author} {\bibfnamefont {E.~M.}\ \bibnamefont
  {Standish}},\ }\href {\doibase 10.1103/PhysRevLett.61.1159} {\bibfield
  {journal} {\bibinfo  {journal} {Phys. Rev. Lett.}\ }\textbf {\bibinfo
  {volume} {61}},\ \bibinfo {pages} {1159} (\bibinfo {year}
  {1988})}\BibitemShut {NoStop}%
\bibitem [{\citenamefont {Tsai}\ \emph {et~al.}(2021)\citenamefont {Tsai},
  \citenamefont {Wu}, \citenamefont {Vagnozzi},\ and\ \citenamefont
  {Visinelli}}]{Tsai:2021irw}%
  \BibitemOpen
  \bibfield  {author} {\bibinfo {author} {\bibfnamefont {Y.-D.}\ \bibnamefont
  {Tsai}}, \bibinfo {author} {\bibfnamefont {Y.}~\bibnamefont {Wu}}, \bibinfo
  {author} {\bibfnamefont {S.}~\bibnamefont {Vagnozzi}}, \ and\ \bibinfo
  {author} {\bibfnamefont {L.}~\bibnamefont {Visinelli}},\ }\href@noop {} {\
  (\bibinfo {year} {2021})},\ \Eprint {http://arxiv.org/abs/2107.04038}
  {arXiv:2107.04038 [hep-ph]} \BibitemShut {NoStop}%
\bibitem [{\citenamefont {Cardone}\ and\ \citenamefont
  {Capozziello}(2011)}]{Cardone:2011ze}%
  \BibitemOpen
  \bibfield  {author} {\bibinfo {author} {\bibfnamefont {V.~F.}\ \bibnamefont
  {Cardone}}\ and\ \bibinfo {author} {\bibfnamefont {S.}~\bibnamefont
  {Capozziello}},\ }\href {\doibase 10.1111/j.1365-2966.2011.18465.x}
  {\bibfield  {journal} {\bibinfo  {journal} {Mon. Not. Roy. Astron. Soc.}\
  }\textbf {\bibinfo {volume} {414}},\ \bibinfo {pages} {1301} (\bibinfo {year}
  {2011})},\ \Eprint {http://arxiv.org/abs/1102.0916} {arXiv:1102.0916
  [astro-ph.CO]} \BibitemShut {NoStop}%
\bibitem [{\citenamefont {Fischbach}\ and\ \citenamefont
  {Talmadge}(1992)}]{Fischbach:1992fa}%
  \BibitemOpen
  \bibfield  {author} {\bibinfo {author} {\bibfnamefont {E.}~\bibnamefont
  {Fischbach}}\ and\ \bibinfo {author} {\bibfnamefont {C.}~\bibnamefont
  {Talmadge}},\ }\href {\doibase 10.1038/356207a0} {\bibfield  {journal}
  {\bibinfo  {journal} {Nature}\ }\textbf {\bibinfo {volume} {356}},\ \bibinfo
  {pages} {207} (\bibinfo {year} {1992})}\BibitemShut {NoStop}%
\bibitem [{\citenamefont {Raffelt}(1999)}]{Raffelt:1999tx}%
  \BibitemOpen
  \bibfield  {author} {\bibinfo {author} {\bibfnamefont {G.~G.}\ \bibnamefont
  {Raffelt}},\ }\href {\doibase 10.1146/annurev.nucl.49.1.163} {\bibfield
  {journal} {\bibinfo  {journal} {Ann. Rev. Nucl. Part. Sci.}\ }\textbf
  {\bibinfo {volume} {49}},\ \bibinfo {pages} {163} (\bibinfo {year} {1999})},\
  \Eprint {http://arxiv.org/abs/hep-ph/9903472} {arXiv:hep-ph/9903472}
  \BibitemShut {NoStop}%
\bibitem [{\citenamefont {Adelberger}\ \emph {et~al.}(2007)\citenamefont
  {Adelberger}, \citenamefont {Heckel}, \citenamefont {Hoedl}, \citenamefont
  {Hoyle}, \citenamefont {Kapner},\ and\ \citenamefont
  {Upadhye}}]{Adelberger:2006dh}%
  \BibitemOpen
  \bibfield  {author} {\bibinfo {author} {\bibfnamefont {E.~G.}\ \bibnamefont
  {Adelberger}}, \bibinfo {author} {\bibfnamefont {B.~R.}\ \bibnamefont
  {Heckel}}, \bibinfo {author} {\bibfnamefont {S.~A.}\ \bibnamefont {Hoedl}},
  \bibinfo {author} {\bibfnamefont {C.~D.}\ \bibnamefont {Hoyle}}, \bibinfo
  {author} {\bibfnamefont {D.~J.}\ \bibnamefont {Kapner}}, \ and\ \bibinfo
  {author} {\bibfnamefont {A.}~\bibnamefont {Upadhye}},\ }\href {\doibase
  10.1103/PhysRevLett.98.131104} {\bibfield  {journal} {\bibinfo  {journal}
  {Phys. Rev. Lett.}\ }\textbf {\bibinfo {volume} {98}},\ \bibinfo {pages}
  {131104} (\bibinfo {year} {2007})},\ \Eprint
  {http://arxiv.org/abs/hep-ph/0611223} {arXiv:hep-ph/0611223} \BibitemShut
  {NoStop}%
\bibitem [{\citenamefont {Niebauer}\ \emph {et~al.}(1987)\citenamefont
  {Niebauer}, \citenamefont {Mchugh},\ and\ \citenamefont
  {Faller}}]{Niebauer:1987ua}%
  \BibitemOpen
  \bibfield  {author} {\bibinfo {author} {\bibfnamefont {T.~M.}\ \bibnamefont
  {Niebauer}}, \bibinfo {author} {\bibfnamefont {M.~P.}\ \bibnamefont
  {Mchugh}}, \ and\ \bibinfo {author} {\bibfnamefont {J.~E.}\ \bibnamefont
  {Faller}},\ }\href {\doibase 10.1103/PhysRevLett.59.609} {\bibfield
  {journal} {\bibinfo  {journal} {Phys. Rev. Lett.}\ }\textbf {\bibinfo
  {volume} {59}},\ \bibinfo {pages} {609} (\bibinfo {year} {1987})}\BibitemShut
  {NoStop}%
\bibitem [{\citenamefont {Hoyle}\ \emph {et~al.}(2001)\citenamefont {Hoyle},
  \citenamefont {Schmidt}, \citenamefont {Heckel}, \citenamefont {Adelberger},
  \citenamefont {Gundlach}, \citenamefont {Kapner},\ and\ \citenamefont
  {Swanson}}]{Hoyle:2000cv}%
  \BibitemOpen
  \bibfield  {author} {\bibinfo {author} {\bibfnamefont {C.~D.}\ \bibnamefont
  {Hoyle}}, \bibinfo {author} {\bibfnamefont {U.}~\bibnamefont {Schmidt}},
  \bibinfo {author} {\bibfnamefont {B.~R.}\ \bibnamefont {Heckel}}, \bibinfo
  {author} {\bibfnamefont {E.~G.}\ \bibnamefont {Adelberger}}, \bibinfo
  {author} {\bibfnamefont {J.~H.}\ \bibnamefont {Gundlach}}, \bibinfo {author}
  {\bibfnamefont {D.~J.}\ \bibnamefont {Kapner}}, \ and\ \bibinfo {author}
  {\bibfnamefont {H.~E.}\ \bibnamefont {Swanson}},\ }\href {\doibase
  10.1103/PhysRevLett.86.1418} {\bibfield  {journal} {\bibinfo  {journal}
  {Phys. Rev. Lett.}\ }\textbf {\bibinfo {volume} {86}},\ \bibinfo {pages}
  {1418} (\bibinfo {year} {2001})},\ \Eprint
  {http://arxiv.org/abs/hep-ph/0011014} {arXiv:hep-ph/0011014} \BibitemShut
  {NoStop}%
\bibitem [{\citenamefont {Adelberger}\ \emph {et~al.}(2003)\citenamefont
  {Adelberger}, \citenamefont {Heckel},\ and\ \citenamefont
  {Nelson}}]{Adelberger:2003zx}%
  \BibitemOpen
  \bibfield  {author} {\bibinfo {author} {\bibfnamefont {E.~G.}\ \bibnamefont
  {Adelberger}}, \bibinfo {author} {\bibfnamefont {B.~R.}\ \bibnamefont
  {Heckel}}, \ and\ \bibinfo {author} {\bibfnamefont {A.~E.}\ \bibnamefont
  {Nelson}},\ }\href {\doibase 10.1146/annurev.nucl.53.041002.110503}
  {\bibfield  {journal} {\bibinfo  {journal} {Ann. Rev. Nucl. Part. Sci.}\
  }\textbf {\bibinfo {volume} {53}},\ \bibinfo {pages} {77} (\bibinfo {year}
  {2003})},\ \Eprint {http://arxiv.org/abs/hep-ph/0307284}
  {arXiv:hep-ph/0307284} \BibitemShut {NoStop}%
\bibitem [{\citenamefont {Peron}(2014)}]{Peron:2014pba}%
  \BibitemOpen
  \bibfield  {author} {\bibinfo {author} {\bibfnamefont {R.}~\bibnamefont
  {Peron}},\ }\href {\doibase 10.1155/2014/791367} {\bibfield  {journal}
  {\bibinfo  {journal} {Adv. High Energy Phys.}\ }\textbf {\bibinfo {volume}
  {2014}},\ \bibinfo {pages} {791367} (\bibinfo {year} {2014})}\BibitemShut
  {NoStop}%
\bibitem [{\citenamefont {Williams}\ \emph {et~al.}(2009)\citenamefont
  {Williams}, \citenamefont {Turyshev},\ and\ \citenamefont
  {Boggs}}]{Williams:2005rv}%
  \BibitemOpen
  \bibfield  {author} {\bibinfo {author} {\bibfnamefont {J.~G.}\ \bibnamefont
  {Williams}}, \bibinfo {author} {\bibfnamefont {S.~G.}\ \bibnamefont
  {Turyshev}}, \ and\ \bibinfo {author} {\bibfnamefont {D.~H.}\ \bibnamefont
  {Boggs}},\ }\href {\doibase 10.1142/S021827180901500X} {\bibfield  {journal}
  {\bibinfo  {journal} {Int. J. Mod. Phys. D}\ }\textbf {\bibinfo {volume}
  {18}},\ \bibinfo {pages} {1129} (\bibinfo {year} {2009})},\ \Eprint
  {http://arxiv.org/abs/gr-qc/0507083} {arXiv:gr-qc/0507083} \BibitemShut
  {NoStop}%
\bibitem [{\citenamefont {Muller}\ \emph {et~al.}(2008)\citenamefont {Muller},
  \citenamefont {Williams},\ and\ \citenamefont {Turyshev}}]{Muller:2005sr}%
  \BibitemOpen
  \bibfield  {author} {\bibinfo {author} {\bibfnamefont {J.}~\bibnamefont
  {Muller}}, \bibinfo {author} {\bibfnamefont {J.~G.}\ \bibnamefont
  {Williams}}, \ and\ \bibinfo {author} {\bibfnamefont {S.~G.}\ \bibnamefont
  {Turyshev}},\ }\href {\doibase 10.1007/978-3-540-34377-6_21} {\bibfield
  {journal} {\bibinfo  {journal} {Astrophys. Space Sci. Libr.}\ }\textbf
  {\bibinfo {volume} {349}},\ \bibinfo {pages} {457} (\bibinfo {year}
  {2008})},\ \Eprint {http://arxiv.org/abs/gr-qc/0509114} {arXiv:gr-qc/0509114}
  \BibitemShut {NoStop}%
\bibitem [{\citenamefont {Konopliv}\ \emph {et~al.}(2011)\citenamefont
  {Konopliv}, \citenamefont {Asmar}, \citenamefont {Folkner}, \citenamefont
  {Karatekin}, \citenamefont {Nunes}, \citenamefont {Smrekar}, \citenamefont
  {Yoder},\ and\ \citenamefont {Zuber}}]{KONOPLIV2011401}%
  \BibitemOpen
  \bibfield  {author} {\bibinfo {author} {\bibfnamefont {A.~S.}\ \bibnamefont
  {Konopliv}}, \bibinfo {author} {\bibfnamefont {S.~W.}\ \bibnamefont {Asmar}},
  \bibinfo {author} {\bibfnamefont {W.~M.}\ \bibnamefont {Folkner}}, \bibinfo
  {author} {\bibfnamefont {O.}~\bibnamefont {Karatekin}}, \bibinfo {author}
  {\bibfnamefont {D.~C.}\ \bibnamefont {Nunes}}, \bibinfo {author}
  {\bibfnamefont {S.~E.}\ \bibnamefont {Smrekar}}, \bibinfo {author}
  {\bibfnamefont {C.~F.}\ \bibnamefont {Yoder}}, \ and\ \bibinfo {author}
  {\bibfnamefont {M.~T.}\ \bibnamefont {Zuber}},\ }\href {\doibase
  https://doi.org/10.1016/j.icarus.2010.10.004} {\bibfield  {journal} {\bibinfo
   {journal} {Icarus}\ }\textbf {\bibinfo {volume} {211}},\ \bibinfo {pages}
  {401} (\bibinfo {year} {2011})}\BibitemShut {NoStop}%
\bibitem [{\citenamefont {Hees}\ \emph {et~al.}(2014)\citenamefont {Hees},
  \citenamefont {Folkner}, \citenamefont {Jacobson},\ and\ \citenamefont
  {Park}}]{Hees:2014kta}%
  \BibitemOpen
  \bibfield  {author} {\bibinfo {author} {\bibfnamefont {A.}~\bibnamefont
  {Hees}}, \bibinfo {author} {\bibfnamefont {W.~M.}\ \bibnamefont {Folkner}},
  \bibinfo {author} {\bibfnamefont {R.~A.}\ \bibnamefont {Jacobson}}, \ and\
  \bibinfo {author} {\bibfnamefont {R.~S.}\ \bibnamefont {Park}},\ }\href
  {\doibase 10.1103/PhysRevD.89.102002} {\bibfield  {journal} {\bibinfo
  {journal} {Phys. Rev. D}\ }\textbf {\bibinfo {volume} {89}},\ \bibinfo
  {pages} {102002} (\bibinfo {year} {2014})},\ \Eprint
  {http://arxiv.org/abs/1402.6950} {arXiv:1402.6950 [gr-qc]} \BibitemShut
  {NoStop}%
\bibitem [{\citenamefont {Li}\ \emph {et~al.}(2014)\citenamefont {Li},
  \citenamefont {Yuan}, \citenamefont {Lu},\ and\ \citenamefont
  {Xie}}]{Li:2014hya}%
  \BibitemOpen
  \bibfield  {author} {\bibinfo {author} {\bibfnamefont {Z.-W.}\ \bibnamefont
  {Li}}, \bibinfo {author} {\bibfnamefont {S.-F.}\ \bibnamefont {Yuan}},
  \bibinfo {author} {\bibfnamefont {C.}~\bibnamefont {Lu}}, \ and\ \bibinfo
  {author} {\bibfnamefont {Y.}~\bibnamefont {Xie}},\ }\href {\doibase
  10.1088/1674-4527/14/2/002} {\bibfield  {journal} {\bibinfo  {journal} {Res.
  Astron. Astrophys.}\ }\textbf {\bibinfo {volume} {14}},\ \bibinfo {pages}
  {139} (\bibinfo {year} {2014})}\BibitemShut {NoStop}%
\bibitem [{\citenamefont {Benisty}(2022)}]{Benisty:2022txp}%
  \BibitemOpen
  \bibfield  {author} {\bibinfo {author} {\bibfnamefont {D.}~\bibnamefont
  {Benisty}},\ }\href {\doibase 10.1103/PhysRevD.106.043001} {\bibfield
  {journal} {\bibinfo  {journal} {Phys. Rev. D}\ }\textbf {\bibinfo {volume}
  {106}},\ \bibinfo {pages} {043001} (\bibinfo {year} {2022})},\ \Eprint
  {http://arxiv.org/abs/2207.08235} {arXiv:2207.08235 [gr-qc]} \BibitemShut
  {NoStop}%
\bibitem [{\citenamefont {Gillessen}\ \emph {et~al.}(2009)\citenamefont
  {Gillessen}, \citenamefont {Eisenhauer}, \citenamefont {Trippe},
  \citenamefont {Alexander}, \citenamefont {Genzel}, \citenamefont {Martins},\
  and\ \citenamefont {Ott}}]{Gillessen:2008qv}%
  \BibitemOpen
  \bibfield  {author} {\bibinfo {author} {\bibfnamefont {S.}~\bibnamefont
  {Gillessen}}, \bibinfo {author} {\bibfnamefont {F.}~\bibnamefont
  {Eisenhauer}}, \bibinfo {author} {\bibfnamefont {S.}~\bibnamefont {Trippe}},
  \bibinfo {author} {\bibfnamefont {T.}~\bibnamefont {Alexander}}, \bibinfo
  {author} {\bibfnamefont {R.}~\bibnamefont {Genzel}}, \bibinfo {author}
  {\bibfnamefont {F.}~\bibnamefont {Martins}}, \ and\ \bibinfo {author}
  {\bibfnamefont {T.}~\bibnamefont {Ott}},\ }\href {\doibase
  10.1088/0004-637X/692/2/1075} {\bibfield  {journal} {\bibinfo  {journal}
  {Astrophys. J.}\ }\textbf {\bibinfo {volume} {692}},\ \bibinfo {pages} {1075}
  (\bibinfo {year} {2009})},\ \Eprint {http://arxiv.org/abs/0810.4674}
  {arXiv:0810.4674 [astro-ph]} \BibitemShut {NoStop}%
\bibitem [{\citenamefont {Akiyama}\ \emph {et~al.}(2022)\citenamefont {Akiyama}
  \emph {et~al.}}]{EventHorizonTelescope:2022xnr}%
  \BibitemOpen
  \bibfield  {author} {\bibinfo {author} {\bibfnamefont {K.}~\bibnamefont
  {Akiyama}} \emph {et~al.} (\bibinfo {collaboration} {Event Horizon
  Telescope}),\ }\href {\doibase 10.3847/2041-8213/ac6674} {\bibfield
  {journal} {\bibinfo  {journal} {Astrophys. J. Lett.}\ }\textbf {\bibinfo
  {volume} {930}},\ \bibinfo {pages} {L12} (\bibinfo {year}
  {2022})}\BibitemShut {NoStop}%
\bibitem [{\citenamefont {Hees}\ \emph {et~al.}(2017)\citenamefont {Hees} \emph
  {et~al.}}]{Hees:2017aal}%
  \BibitemOpen
  \bibfield  {author} {\bibinfo {author} {\bibfnamefont {A.}~\bibnamefont
  {Hees}} \emph {et~al.},\ }\href {\doibase 10.1103/PhysRevLett.118.211101}
  {\bibfield  {journal} {\bibinfo  {journal} {Phys. Rev. Lett.}\ }\textbf
  {\bibinfo {volume} {118}},\ \bibinfo {pages} {211101} (\bibinfo {year}
  {2017})},\ \Eprint {http://arxiv.org/abs/1705.07902} {arXiv:1705.07902
  [astro-ph.GA]} \BibitemShut {NoStop}%
\bibitem [{\citenamefont {Bernus}\ \emph {et~al.}(2019)\citenamefont {Bernus},
  \citenamefont {Minazzoli}, \citenamefont {Fienga}, \citenamefont {Gastineau},
  \citenamefont {Laskar},\ and\ \citenamefont {Deram}}]{Bernus:2019rgl}%
  \BibitemOpen
  \bibfield  {author} {\bibinfo {author} {\bibfnamefont {L.}~\bibnamefont
  {Bernus}}, \bibinfo {author} {\bibfnamefont {O.}~\bibnamefont {Minazzoli}},
  \bibinfo {author} {\bibfnamefont {A.}~\bibnamefont {Fienga}}, \bibinfo
  {author} {\bibfnamefont {M.}~\bibnamefont {Gastineau}}, \bibinfo {author}
  {\bibfnamefont {J.}~\bibnamefont {Laskar}}, \ and\ \bibinfo {author}
  {\bibfnamefont {P.}~\bibnamefont {Deram}},\ }\href {\doibase
  10.1103/PhysRevLett.123.161103} {\bibfield  {journal} {\bibinfo  {journal}
  {Phys. Rev. Lett.}\ }\textbf {\bibinfo {volume} {123}},\ \bibinfo {pages}
  {161103} (\bibinfo {year} {2019})},\ \Eprint
  {http://arxiv.org/abs/1901.04307} {arXiv:1901.04307 [gr-qc]} \BibitemShut
  {NoStop}%
\bibitem [{\citenamefont {Will}(2018)}]{Will:2018gku}%
  \BibitemOpen
  \bibfield  {author} {\bibinfo {author} {\bibfnamefont {C.~M.}\ \bibnamefont
  {Will}},\ }\href {\doibase 10.1088/1361-6382/aad13c} {\bibfield  {journal}
  {\bibinfo  {journal} {Class. Quant. Grav.}\ }\textbf {\bibinfo {volume}
  {35}},\ \bibinfo {pages} {17LT01} (\bibinfo {year} {2018})},\ \Eprint
  {http://arxiv.org/abs/1805.10523} {arXiv:1805.10523 [gr-qc]} \BibitemShut
  {NoStop}%
\bibitem [{\citenamefont {Jovanovi\'c}\ \emph {et~al.}(2021)\citenamefont
  {Jovanovi\'c}, \citenamefont {Borka}, \citenamefont {Borka~Jovanovi\'c},\
  and\ \citenamefont {Zakharov}}]{Jovanovic:2021hrz}%
  \BibitemOpen
  \bibfield  {author} {\bibinfo {author} {\bibfnamefont {P.}~\bibnamefont
  {Jovanovi\'c}}, \bibinfo {author} {\bibfnamefont {D.}~\bibnamefont {Borka}},
  \bibinfo {author} {\bibfnamefont {V.}~\bibnamefont {Borka~Jovanovi\'c}}, \
  and\ \bibinfo {author} {\bibfnamefont {A.~F.}\ \bibnamefont {Zakharov}},\
  }\href {\doibase 10.1140/epjd/s10053-021-00154-z} {\bibfield  {journal}
  {\bibinfo  {journal} {Eur. Phys. J. D}\ }\textbf {\bibinfo {volume} {75}},\
  \bibinfo {pages} {145} (\bibinfo {year} {2021})},\ \Eprint
  {http://arxiv.org/abs/2105.03403} {arXiv:2105.03403 [astro-ph.GA]}
  \BibitemShut {NoStop}%
\bibitem [{\citenamefont {de~Rham}\ \emph {et~al.}(2017)\citenamefont
  {de~Rham}, \citenamefont {Deskins}, \citenamefont {Tolley},\ and\
  \citenamefont {Zhou}}]{deRham:2016nuf}%
  \BibitemOpen
  \bibfield  {author} {\bibinfo {author} {\bibfnamefont {C.}~\bibnamefont
  {de~Rham}}, \bibinfo {author} {\bibfnamefont {J.~T.}\ \bibnamefont
  {Deskins}}, \bibinfo {author} {\bibfnamefont {A.~J.}\ \bibnamefont {Tolley}},
  \ and\ \bibinfo {author} {\bibfnamefont {S.-Y.}\ \bibnamefont {Zhou}},\
  }\href {\doibase 10.1103/RevModPhys.89.025004} {\bibfield  {journal}
  {\bibinfo  {journal} {Rev. Mod. Phys.}\ }\textbf {\bibinfo {volume} {89}},\
  \bibinfo {pages} {025004} (\bibinfo {year} {2017})},\ \Eprint
  {http://arxiv.org/abs/1606.08462} {arXiv:1606.08462 [astro-ph.CO]}
  \BibitemShut {NoStop}%
\bibitem [{\citenamefont {Bernus}\ \emph {et~al.}(2020)\citenamefont {Bernus},
  \citenamefont {Minazzoli}, \citenamefont {Fienga}, \citenamefont {Gastineau},
  \citenamefont {Laskar}, \citenamefont {Deram},\ and\ \citenamefont
  {Di~Ruscio}}]{Bernus_2020}%
  \BibitemOpen
  \bibfield  {author} {\bibinfo {author} {\bibfnamefont {L.}~\bibnamefont
  {Bernus}}, \bibinfo {author} {\bibfnamefont {O.}~\bibnamefont {Minazzoli}},
  \bibinfo {author} {\bibfnamefont {A.}~\bibnamefont {Fienga}}, \bibinfo
  {author} {\bibfnamefont {M.}~\bibnamefont {Gastineau}}, \bibinfo {author}
  {\bibfnamefont {J.}~\bibnamefont {Laskar}}, \bibinfo {author} {\bibfnamefont
  {P.}~\bibnamefont {Deram}}, \ and\ \bibinfo {author} {\bibfnamefont
  {A.}~\bibnamefont {Di~Ruscio}},\ }\href {\doibase
  10.1103/PhysRevD.102.021501} {\bibfield  {journal} {\bibinfo  {journal}
  {Phys. Rev. D}\ }\textbf {\bibinfo {volume} {102}},\ \bibinfo {pages}
  {021501} (\bibinfo {year} {2020})},\ \Eprint
  {http://arxiv.org/abs/2006.12304} {arXiv:2006.12304 [gr-qc]} \BibitemShut
  {NoStop}%
\bibitem [{\citenamefont {Iorio}(2007)}]{Iorio:2007gq}%
  \BibitemOpen
  \bibfield  {author} {\bibinfo {author} {\bibfnamefont {L.}~\bibnamefont
  {Iorio}},\ }\href {\doibase 10.1088/1126-6708/2007/10/041} {\bibfield
  {journal} {\bibinfo  {journal} {JHEP}\ }\textbf {\bibinfo {volume} {10}},\
  \bibinfo {pages} {041} (\bibinfo {year} {2007})},\ \Eprint
  {http://arxiv.org/abs/0708.1080} {arXiv:0708.1080 [gr-qc]} \BibitemShut
  {NoStop}%
\bibitem [{\citenamefont {Finn}\ and\ \citenamefont
  {Sutton}(2002)}]{Finn:2001qi}%
  \BibitemOpen
  \bibfield  {author} {\bibinfo {author} {\bibfnamefont {L.~S.}\ \bibnamefont
  {Finn}}\ and\ \bibinfo {author} {\bibfnamefont {P.~J.}\ \bibnamefont
  {Sutton}},\ }\href {\doibase 10.1103/PhysRevD.65.044022} {\bibfield
  {journal} {\bibinfo  {journal} {Phys. Rev. D}\ }\textbf {\bibinfo {volume}
  {65}},\ \bibinfo {pages} {044022} (\bibinfo {year} {2002})},\ \Eprint
  {http://arxiv.org/abs/gr-qc/0109049} {arXiv:gr-qc/0109049} \BibitemShut
  {NoStop}%
\bibitem [{\citenamefont {Miao}\ \emph {et~al.}(2019)\citenamefont {Miao},
  \citenamefont {Shao},\ and\ \citenamefont {Ma}}]{Miao:2019nhf}%
  \BibitemOpen
  \bibfield  {author} {\bibinfo {author} {\bibfnamefont {X.}~\bibnamefont
  {Miao}}, \bibinfo {author} {\bibfnamefont {L.}~\bibnamefont {Shao}}, \ and\
  \bibinfo {author} {\bibfnamefont {B.-Q.}\ \bibnamefont {Ma}},\ }\href
  {\doibase 10.1103/PhysRevD.99.123015} {\bibfield  {journal} {\bibinfo
  {journal} {Phys. Rev. D}\ }\textbf {\bibinfo {volume} {99}},\ \bibinfo
  {pages} {123015} (\bibinfo {year} {2019})},\ \Eprint
  {http://arxiv.org/abs/1905.12836} {arXiv:1905.12836 [astro-ph.CO]}
  \BibitemShut {NoStop}%
\bibitem [{\citenamefont {de~Rham}\ \emph {et~al.}(2013)\citenamefont
  {de~Rham}, \citenamefont {Tolley},\ and\ \citenamefont
  {Wesley}}]{deRham:2012fw}%
  \BibitemOpen
  \bibfield  {author} {\bibinfo {author} {\bibfnamefont {C.}~\bibnamefont
  {de~Rham}}, \bibinfo {author} {\bibfnamefont {A.~J.}\ \bibnamefont {Tolley}},
  \ and\ \bibinfo {author} {\bibfnamefont {D.~H.}\ \bibnamefont {Wesley}},\
  }\href {\doibase 10.1103/PhysRevD.87.044025} {\bibfield  {journal} {\bibinfo
  {journal} {Phys. Rev. D}\ }\textbf {\bibinfo {volume} {87}},\ \bibinfo
  {pages} {044025} (\bibinfo {year} {2013})},\ \Eprint
  {http://arxiv.org/abs/1208.0580} {arXiv:1208.0580 [gr-qc]} \BibitemShut
  {NoStop}%
\bibitem [{\citenamefont {Shao}\ \emph {et~al.}(2020)\citenamefont {Shao},
  \citenamefont {Wex},\ and\ \citenamefont {Zhou}}]{Shao:2020fka}%
  \BibitemOpen
  \bibfield  {author} {\bibinfo {author} {\bibfnamefont {L.}~\bibnamefont
  {Shao}}, \bibinfo {author} {\bibfnamefont {N.}~\bibnamefont {Wex}}, \ and\
  \bibinfo {author} {\bibfnamefont {S.-Y.}\ \bibnamefont {Zhou}},\ }\href
  {\doibase 10.1103/PhysRevD.102.024069} {\bibfield  {journal} {\bibinfo
  {journal} {Phys. Rev. D}\ }\textbf {\bibinfo {volume} {102}},\ \bibinfo
  {pages} {024069} (\bibinfo {year} {2020})},\ \Eprint
  {http://arxiv.org/abs/2007.04531} {arXiv:2007.04531 [gr-qc]} \BibitemShut
  {NoStop}%
\bibitem [{\citenamefont {Will}(1998)}]{Will:1997bb}%
  \BibitemOpen
  \bibfield  {author} {\bibinfo {author} {\bibfnamefont {C.~M.}\ \bibnamefont
  {Will}},\ }\href {\doibase 10.1103/PhysRevD.57.2061} {\bibfield  {journal}
  {\bibinfo  {journal} {Phys. Rev. D}\ }\textbf {\bibinfo {volume} {57}},\
  \bibinfo {pages} {2061} (\bibinfo {year} {1998})},\ \Eprint
  {http://arxiv.org/abs/gr-qc/9709011} {arXiv:gr-qc/9709011} \BibitemShut
  {NoStop}%
\bibitem [{\citenamefont {Abbott}\ \emph {et~al.}(2016)\citenamefont {Abbott}
  \emph {et~al.}}]{LIGOScientific:2016lio}%
  \BibitemOpen
  \bibfield  {author} {\bibinfo {author} {\bibfnamefont {B.~P.}\ \bibnamefont
  {Abbott}} \emph {et~al.} (\bibinfo {collaboration} {LIGO Scientific,
  Virgo}),\ }\href {\doibase 10.1103/PhysRevLett.116.221101} {\bibfield
  {journal} {\bibinfo  {journal} {Phys. Rev. Lett.}\ }\textbf {\bibinfo
  {volume} {116}},\ \bibinfo {pages} {221101} (\bibinfo {year} {2016})},\
  \bibinfo {note} {[Erratum: Phys.Rev.Lett. 121, 129902 (2018)]},\ \Eprint
  {http://arxiv.org/abs/1602.03841} {arXiv:1602.03841 [gr-qc]} \BibitemShut
  {NoStop}%
\bibitem [{\citenamefont {Abbott}\ \emph {et~al.}(2021)\citenamefont {Abbott}
  \emph {et~al.}}]{LIGOScientific:2021sio}%
  \BibitemOpen
  \bibfield  {author} {\bibinfo {author} {\bibfnamefont {R.}~\bibnamefont
  {Abbott}} \emph {et~al.} (\bibinfo {collaboration} {LIGO Scientific, VIRGO,
  KAGRA}),\ }\href@noop {} {\  (\bibinfo {year} {2021})},\ \Eprint
  {http://arxiv.org/abs/2112.06861} {arXiv:2112.06861 [gr-qc]} \BibitemShut
  {NoStop}%
\bibitem [{\citenamefont
  {Pi\'orkowska-Kurpas}(2022)}]{Piorkowska-Kurpas:2022xmb}%
  \BibitemOpen
  \bibfield  {author} {\bibinfo {author} {\bibfnamefont {A.}~\bibnamefont
  {Pi\'orkowska-Kurpas}},\ }\href {\doibase 10.3390/universe8020083} {\bibfield
   {journal} {\bibinfo  {journal} {Universe}\ }\textbf {\bibinfo {volume}
  {8}},\ \bibinfo {pages} {83} (\bibinfo {year} {2022})}\BibitemShut {NoStop}%
\bibitem [{\citenamefont {Taylor}(1992)}]{Taylor:1992kea}%
  \BibitemOpen
  \bibfield  {author} {\bibinfo {author} {\bibfnamefont {J.~H.}\ \bibnamefont
  {Taylor}},\ }\href {\doibase 10.1098/rsta.1992.0088} {\bibfield  {journal}
  {\bibinfo  {journal} {Phil. Trans. A. Math. Phys. Eng. Sci.}\ }\textbf
  {\bibinfo {volume} {341}},\ \bibinfo {pages} {117} (\bibinfo {year}
  {1992})}\BibitemShut {NoStop}%
\bibitem [{\citenamefont {Shao}\ and\ \citenamefont
  {Wex}(2016)}]{Shao:2016ezh}%
  \BibitemOpen
  \bibfield  {author} {\bibinfo {author} {\bibfnamefont {L.}~\bibnamefont
  {Shao}}\ and\ \bibinfo {author} {\bibfnamefont {N.}~\bibnamefont {Wex}},\
  }\href {\doibase 10.1007/s11433-016-0087-6} {\bibfield  {journal} {\bibinfo
  {journal} {Sci. China Phys. Mech. Astron.}\ }\textbf {\bibinfo {volume}
  {59}},\ \bibinfo {pages} {699501} (\bibinfo {year} {2016})},\ \Eprint
  {http://arxiv.org/abs/1604.03662} {arXiv:1604.03662 [gr-qc]} \BibitemShut
  {NoStop}%
\bibitem [{\citenamefont {Wex}(2014)}]{Wex:2014nva}%
  \BibitemOpen
  \bibfield  {author} {\bibinfo {author} {\bibfnamefont {N.}~\bibnamefont
  {Wex}},\ }in\ \href@noop {} {\emph {\bibinfo {booktitle} {{Frontiers in
  Relativistic Celestial Mechanics: Applications and Experiments}}}},\
  Vol.~\bibinfo {volume} {2},\ \bibinfo {editor} {edited by\ \bibinfo {editor}
  {\bibfnamefont {S.~M.}\ \bibnamefont {Kopeikin}}}\ (\bibinfo  {publisher}
  {Walter de Gruyter GmbH, Berlin/Boston},\ \bibinfo {year} {2014})\
  p.~\bibinfo {pages} {39},\ \Eprint {http://arxiv.org/abs/1402.5594}
  {arXiv:1402.5594 [gr-qc]} \BibitemShut {NoStop}%
\bibitem [{\citenamefont {Kramer}(2016)}]{Kramer:2016kwa}%
  \BibitemOpen
  \bibfield  {author} {\bibinfo {author} {\bibfnamefont {M.}~\bibnamefont
  {Kramer}},\ }\href {\doibase 10.1142/S0218271816300299} {\bibfield  {journal}
  {\bibinfo  {journal} {Int. J. Mod. Phys. D}\ }\textbf {\bibinfo {volume}
  {25}},\ \bibinfo {pages} {1630029} (\bibinfo {year} {2016})},\ \Eprint
  {http://arxiv.org/abs/1606.03843} {arXiv:1606.03843 [astro-ph.HE]}
  \BibitemShut {NoStop}%
\bibitem [{\citenamefont {Miao}\ \emph {et~al.}(2020)\citenamefont {Miao},
  \citenamefont {Zhao}, \citenamefont {Shao}, \citenamefont {Wex},
  \citenamefont {Kramer},\ and\ \citenamefont {Ma}}]{Miao:2020wph}%
  \BibitemOpen
  \bibfield  {author} {\bibinfo {author} {\bibfnamefont {X.}~\bibnamefont
  {Miao}}, \bibinfo {author} {\bibfnamefont {J.}~\bibnamefont {Zhao}}, \bibinfo
  {author} {\bibfnamefont {L.}~\bibnamefont {Shao}}, \bibinfo {author}
  {\bibfnamefont {N.}~\bibnamefont {Wex}}, \bibinfo {author} {\bibfnamefont
  {M.}~\bibnamefont {Kramer}}, \ and\ \bibinfo {author} {\bibfnamefont {B.-Q.}\
  \bibnamefont {Ma}},\ }\href {\doibase 10.3847/1538-4357/ab9dfe} {\bibfield
  {journal} {\bibinfo  {journal} {Astrophys. J.}\ }\textbf {\bibinfo {volume}
  {898}},\ \bibinfo {pages} {69} (\bibinfo {year} {2020})},\ \Eprint
  {http://arxiv.org/abs/2006.09652} {arXiv:2006.09652 [gr-qc]} \BibitemShut
  {NoStop}%
\bibitem [{\citenamefont {Shao}(2022)}]{Shao:2022izp}%
  \BibitemOpen
  \bibfield  {author} {\bibinfo {author} {\bibfnamefont {L.}~\bibnamefont
  {Shao}},\ }\href@noop {} {\  (\bibinfo {year} {2022})},\ \Eprint
  {http://arxiv.org/abs/2206.15187} {arXiv:2206.15187 [gr-qc]} \BibitemShut
  {NoStop}%
\bibitem [{\citenamefont {Shao}\ and\ \citenamefont
  {Yagi}(2022)}]{Shao:2022koz}%
  \BibitemOpen
  \bibfield  {author} {\bibinfo {author} {\bibfnamefont {L.}~\bibnamefont
  {Shao}}\ and\ \bibinfo {author} {\bibfnamefont {K.}~\bibnamefont {Yagi}},\
  }\href {\doibase 10.1016/j.scib.2022.09.018} {\bibfield  {journal} {\bibinfo
  {journal} {Sci. Bull.}\ }\textbf {\bibinfo {volume} {67}},\ \bibinfo {pages}
  {1946} (\bibinfo {year} {2022})},\ \Eprint {http://arxiv.org/abs/2209.03351}
  {arXiv:2209.03351 [gr-qc]} \BibitemShut {NoStop}%
\bibitem [{\citenamefont {Hulse}\ and\ \citenamefont
  {Taylor}(1975)}]{Hulse:1974eb}%
  \BibitemOpen
  \bibfield  {author} {\bibinfo {author} {\bibfnamefont {R.~A.}\ \bibnamefont
  {Hulse}}\ and\ \bibinfo {author} {\bibfnamefont {J.~H.}\ \bibnamefont
  {Taylor}},\ }\href {\doibase 10.1086/181708} {\bibfield  {journal} {\bibinfo
  {journal} {Astrophys. J. Lett.}\ }\textbf {\bibinfo {volume} {195}},\
  \bibinfo {pages} {L51} (\bibinfo {year} {1975})}\BibitemShut {NoStop}%
\bibitem [{\citenamefont {Taylor}\ \emph {et~al.}(1979)\citenamefont {Taylor},
  \citenamefont {Fowler},\ and\ \citenamefont {McCulloch}}]{Taylor:1979zz}%
  \BibitemOpen
  \bibfield  {author} {\bibinfo {author} {\bibfnamefont {J.~H.}\ \bibnamefont
  {Taylor}}, \bibinfo {author} {\bibfnamefont {L.~A.}\ \bibnamefont {Fowler}},
  \ and\ \bibinfo {author} {\bibfnamefont {P.~M.}\ \bibnamefont {McCulloch}},\
  }\href {\doibase 10.1038/277437a0} {\bibfield  {journal} {\bibinfo  {journal}
  {Nature}\ }\textbf {\bibinfo {volume} {277}},\ \bibinfo {pages} {437}
  (\bibinfo {year} {1979})}\BibitemShut {NoStop}%
\bibitem [{\citenamefont {Kramer}\ \emph {et~al.}(2006)\citenamefont {Kramer}
  \emph {et~al.}}]{Kramer:2006nb}%
  \BibitemOpen
  \bibfield  {author} {\bibinfo {author} {\bibfnamefont {M.}~\bibnamefont
  {Kramer}} \emph {et~al.},\ }\href {\doibase 10.1126/science.1132305}
  {\bibfield  {journal} {\bibinfo  {journal} {Science}\ }\textbf {\bibinfo
  {volume} {314}},\ \bibinfo {pages} {97} (\bibinfo {year} {2006})},\ \Eprint
  {http://arxiv.org/abs/astro-ph/0609417} {arXiv:astro-ph/0609417} \BibitemShut
  {NoStop}%
\bibitem [{\citenamefont {Kramer}\ \emph {et~al.}(2021)\citenamefont {Kramer}
  \emph {et~al.}}]{Kramer:2021jcw}%
  \BibitemOpen
  \bibfield  {author} {\bibinfo {author} {\bibfnamefont {M.}~\bibnamefont
  {Kramer}} \emph {et~al.},\ }\href {\doibase 10.1103/PhysRevX.11.041050}
  {\bibfield  {journal} {\bibinfo  {journal} {Phys. Rev. X}\ }\textbf {\bibinfo
  {volume} {11}},\ \bibinfo {pages} {041050} (\bibinfo {year} {2021})},\
  \Eprint {http://arxiv.org/abs/2112.06795} {arXiv:2112.06795 [astro-ph.HE]}
  \BibitemShut {NoStop}%
\bibitem [{\citenamefont {{Gillessen}}\ \emph {et~al.}(2017)\citenamefont
  {{Gillessen}}, \citenamefont {{Plewa}}, \citenamefont {{Eisenhauer}},
  \citenamefont {{Sari}}, \citenamefont {{Waisberg}}, \citenamefont {{Habibi}},
  \citenamefont {{Pfuhl}}, \citenamefont {{George}}, \citenamefont {{Dexter}},
  \citenamefont {{von Fellenberg}}, \citenamefont {{Ott}},\ and\ \citenamefont
  {{Genzel}}}]{2017ApJ...837...30G}%
  \BibitemOpen
  \bibfield  {author} {\bibinfo {author} {\bibfnamefont {S.}~\bibnamefont
  {{Gillessen}}}, \bibinfo {author} {\bibfnamefont {P.~M.}\ \bibnamefont
  {{Plewa}}}, \bibinfo {author} {\bibfnamefont {F.}~\bibnamefont
  {{Eisenhauer}}}, \bibinfo {author} {\bibfnamefont {R.}~\bibnamefont
  {{Sari}}}, \bibinfo {author} {\bibfnamefont {I.}~\bibnamefont {{Waisberg}}},
  \bibinfo {author} {\bibfnamefont {M.}~\bibnamefont {{Habibi}}}, \bibinfo
  {author} {\bibfnamefont {O.}~\bibnamefont {{Pfuhl}}}, \bibinfo {author}
  {\bibfnamefont {E.}~\bibnamefont {{George}}}, \bibinfo {author}
  {\bibfnamefont {J.}~\bibnamefont {{Dexter}}}, \bibinfo {author}
  {\bibfnamefont {S.}~\bibnamefont {{von Fellenberg}}}, \bibinfo {author}
  {\bibfnamefont {T.}~\bibnamefont {{Ott}}}, \ and\ \bibinfo {author}
  {\bibfnamefont {R.}~\bibnamefont {{Genzel}}},\ }\href {\doibase
  10.3847/1538-4357/aa5c41} {\bibfield  {journal} {\bibinfo  {journal}
  {Astrophys. J.}\ }\textbf {\bibinfo {volume} {837}},\ \bibinfo {eid} {30}
  (\bibinfo {year} {2017})},\ \Eprint {http://arxiv.org/abs/1611.09144}
  {arXiv:1611.09144 [astro-ph.GA]} \BibitemShut {NoStop}%
\bibitem [{\citenamefont {Liu}\ \emph {et~al.}(2021)\citenamefont {Liu} \emph
  {et~al.}}]{Liu:2021ziv}%
  \BibitemOpen
  \bibfield  {author} {\bibinfo {author} {\bibfnamefont {K.}~\bibnamefont
  {Liu}} \emph {et~al.},\ }\href {\doibase 10.3847/1538-4357/abf9a2} {\bibfield
   {journal} {\bibinfo  {journal} {Astrophys. J.}\ }\textbf {\bibinfo {volume}
  {914}},\ \bibinfo {pages} {30} (\bibinfo {year} {2021})},\ \Eprint
  {http://arxiv.org/abs/2104.08986} {arXiv:2104.08986 [astro-ph.HE]}
  \BibitemShut {NoStop}%
\bibitem [{\citenamefont {Torne}\ \emph {et~al.}(2021)\citenamefont {Torne}
  \emph {et~al.}}]{Torne:2021yad}%
  \BibitemOpen
  \bibfield  {author} {\bibinfo {author} {\bibfnamefont {P.}~\bibnamefont
  {Torne}} \emph {et~al.},\ }\href {\doibase 10.1051/0004-6361/202140775}
  {\bibfield  {journal} {\bibinfo  {journal} {Astron. Astrophys.}\ }\textbf
  {\bibinfo {volume} {650}},\ \bibinfo {pages} {A95} (\bibinfo {year}
  {2021})},\ \Eprint {http://arxiv.org/abs/2103.16581} {arXiv:2103.16581
  [astro-ph.HE]} \BibitemShut {NoStop}%
\bibitem [{\citenamefont {Suresh}\ \emph {et~al.}(2022)\citenamefont {Suresh},
  \citenamefont {Cordes}, \citenamefont {Chatterjee}, \citenamefont {Gajjar},
  \citenamefont {Perez}, \citenamefont {Siemion}, \citenamefont {Lebofsky},
  \citenamefont {MacMahon},\ and\ \citenamefont {Ng}}]{Suresh:2022vmf}%
  \BibitemOpen
  \bibfield  {author} {\bibinfo {author} {\bibfnamefont {A.}~\bibnamefont
  {Suresh}}, \bibinfo {author} {\bibfnamefont {J.~M.}\ \bibnamefont {Cordes}},
  \bibinfo {author} {\bibfnamefont {S.}~\bibnamefont {Chatterjee}}, \bibinfo
  {author} {\bibfnamefont {V.}~\bibnamefont {Gajjar}}, \bibinfo {author}
  {\bibfnamefont {K.~I.}\ \bibnamefont {Perez}}, \bibinfo {author}
  {\bibfnamefont {A.~P.~V.}\ \bibnamefont {Siemion}}, \bibinfo {author}
  {\bibfnamefont {M.}~\bibnamefont {Lebofsky}}, \bibinfo {author}
  {\bibfnamefont {D.~H.~E.}\ \bibnamefont {MacMahon}}, \ and\ \bibinfo {author}
  {\bibfnamefont {C.}~\bibnamefont {Ng}},\ }\href {\doibase
  10.3847/1538-4357/ac74c0} {\bibfield  {journal} {\bibinfo  {journal}
  {Astrophys. J.}\ }\textbf {\bibinfo {volume} {933}},\ \bibinfo {pages} {121}
  (\bibinfo {year} {2022})},\ \Eprint {http://arxiv.org/abs/2203.00036}
  {arXiv:2203.00036 [astro-ph.HE]} \BibitemShut {NoStop}%
\bibitem [{\citenamefont {Cordes}\ and\ \citenamefont
  {Lazio}(1997)}]{Cordes:1996bt}%
  \BibitemOpen
  \bibfield  {author} {\bibinfo {author} {\bibfnamefont {J.~M.}\ \bibnamefont
  {Cordes}}\ and\ \bibinfo {author} {\bibfnamefont {T.~J.~W.}\ \bibnamefont
  {Lazio}},\ }\href {\doibase 10.1086/303569} {\bibfield  {journal} {\bibinfo
  {journal} {Astrophys. J.}\ }\textbf {\bibinfo {volume} {475}},\ \bibinfo
  {pages} {557} (\bibinfo {year} {1997})},\ \Eprint
  {http://arxiv.org/abs/astro-ph/9608028} {arXiv:astro-ph/9608028} \BibitemShut
  {NoStop}%
\bibitem [{\citenamefont {Wharton}\ \emph {et~al.}(2012)\citenamefont
  {Wharton}, \citenamefont {Chatterjee}, \citenamefont {Cordes}, \citenamefont
  {Deneva},\ and\ \citenamefont {Lazio}}]{Wharton:2011dv}%
  \BibitemOpen
  \bibfield  {author} {\bibinfo {author} {\bibfnamefont {R.~S.}\ \bibnamefont
  {Wharton}}, \bibinfo {author} {\bibfnamefont {S.}~\bibnamefont {Chatterjee}},
  \bibinfo {author} {\bibfnamefont {J.~M.}\ \bibnamefont {Cordes}}, \bibinfo
  {author} {\bibfnamefont {J.~S.}\ \bibnamefont {Deneva}}, \ and\ \bibinfo
  {author} {\bibfnamefont {T.~J.~W.}\ \bibnamefont {Lazio}},\ }\href {\doibase
  10.1088/0004-637X/753/2/108} {\bibfield  {journal} {\bibinfo  {journal}
  {Astrophys. J.}\ }\textbf {\bibinfo {volume} {753}},\ \bibinfo {pages} {108}
  (\bibinfo {year} {2012})},\ \Eprint {http://arxiv.org/abs/1111.4216}
  {arXiv:1111.4216 [astro-ph.HE]} \BibitemShut {NoStop}%
\bibitem [{\citenamefont {Zhang}\ \emph {et~al.}(2014)\citenamefont {Zhang},
  \citenamefont {Lu},\ and\ \citenamefont {Yu}}]{Zhang:2014kva}%
  \BibitemOpen
  \bibfield  {author} {\bibinfo {author} {\bibfnamefont {F.}~\bibnamefont
  {Zhang}}, \bibinfo {author} {\bibfnamefont {Y.}~\bibnamefont {Lu}}, \ and\
  \bibinfo {author} {\bibfnamefont {Q.}~\bibnamefont {Yu}},\ }\href {\doibase
  10.1088/0004-637X/784/2/106} {\bibfield  {journal} {\bibinfo  {journal}
  {Astrophys. J.}\ }\textbf {\bibinfo {volume} {784}},\ \bibinfo {pages} {106}
  (\bibinfo {year} {2014})},\ \Eprint {http://arxiv.org/abs/1402.2505}
  {arXiv:1402.2505 [astro-ph.GA]} \BibitemShut {NoStop}%
\bibitem [{\citenamefont {Bower}\ \emph {et~al.}(2018)\citenamefont {Bower}
  \emph {et~al.}}]{Bower:2018mta}%
  \BibitemOpen
  \bibfield  {author} {\bibinfo {author} {\bibfnamefont {G.~C.}\ \bibnamefont
  {Bower}} \emph {et~al.},\ }\href@noop {} {\bibfield  {journal} {\bibinfo
  {journal} {ASP Conf. Ser.}\ }\textbf {\bibinfo {volume} {517}},\ \bibinfo
  {pages} {793} (\bibinfo {year} {2018})},\ \Eprint
  {http://arxiv.org/abs/1810.06623} {arXiv:1810.06623 [astro-ph.HE]}
  \BibitemShut {NoStop}%
\bibitem [{\citenamefont {Bower}\ \emph {et~al.}(2019)\citenamefont {Bower}
  \emph {et~al.}}]{2019BAAS...51c.438B}%
  \BibitemOpen
  \bibfield  {author} {\bibinfo {author} {\bibfnamefont {G.~C.}\ \bibnamefont
  {Bower}} \emph {et~al.},\ }\href@noop {} {\bibfield  {journal} {\bibinfo
  {journal} {Bull. Am. Astron. Soc.}\ }\textbf {\bibinfo {volume} {51}},\
  \bibinfo {eid} {438} (\bibinfo {year} {2019})}\BibitemShut {NoStop}%
\bibitem [{\citenamefont {Liu}\ \emph {et~al.}(2012)\citenamefont {Liu},
  \citenamefont {Wex}, \citenamefont {Kramer}, \citenamefont {Cordes},\ and\
  \citenamefont {Lazio}}]{Liu:2011ae}%
  \BibitemOpen
  \bibfield  {author} {\bibinfo {author} {\bibfnamefont {K.}~\bibnamefont
  {Liu}}, \bibinfo {author} {\bibfnamefont {N.}~\bibnamefont {Wex}}, \bibinfo
  {author} {\bibfnamefont {M.}~\bibnamefont {Kramer}}, \bibinfo {author}
  {\bibfnamefont {J.~M.}\ \bibnamefont {Cordes}}, \ and\ \bibinfo {author}
  {\bibfnamefont {T.~J.~W.}\ \bibnamefont {Lazio}},\ }\href {\doibase
  10.1088/0004-637X/747/1/1} {\bibfield  {journal} {\bibinfo  {journal}
  {Astrophys. J.}\ }\textbf {\bibinfo {volume} {747}},\ \bibinfo {pages} {1}
  (\bibinfo {year} {2012})},\ \Eprint {http://arxiv.org/abs/1112.2151}
  {arXiv:1112.2151 [astro-ph.HE]} \BibitemShut {NoStop}%
\bibitem [{\citenamefont {Shao}\ \emph {et~al.}(2015)\citenamefont {Shao},
  \citenamefont {Stairs} \emph {et~al.}}]{Shao:2014wja}%
  \BibitemOpen
  \bibfield  {author} {\bibinfo {author} {\bibfnamefont {L.}~\bibnamefont
  {Shao}}, \bibinfo {author} {\bibfnamefont {I.}~\bibnamefont {Stairs}},  \emph
  {et~al.},\ }\href {\doibase 10.22323/1.215.0042} {\bibfield  {journal}
  {\bibinfo  {journal} {PoS}\ }\textbf {\bibinfo {volume} {AASKA14}},\ \bibinfo
  {pages} {042} (\bibinfo {year} {2015})},\ \Eprint
  {http://arxiv.org/abs/1501.00058} {arXiv:1501.00058 [astro-ph.HE]}
  \BibitemShut {NoStop}%
\bibitem [{\citenamefont {Weltman}\ \emph {et~al.}(2020)\citenamefont {Weltman}
  \emph {et~al.}}]{Weltman:2018zrl}%
  \BibitemOpen
  \bibfield  {author} {\bibinfo {author} {\bibfnamefont {A.}~\bibnamefont
  {Weltman}} \emph {et~al.},\ }\href {\doibase 10.1017/pasa.2019.42} {\bibfield
   {journal} {\bibinfo  {journal} {Publ. Astron. Soc. Austral.}\ }\textbf
  {\bibinfo {volume} {37}},\ \bibinfo {pages} {e002} (\bibinfo {year}
  {2020})},\ \Eprint {http://arxiv.org/abs/1810.02680} {arXiv:1810.02680
  [astro-ph.CO]} \BibitemShut {NoStop}%
\bibitem [{\citenamefont {Einstein}\ \emph {et~al.}(1938)\citenamefont
  {Einstein}, \citenamefont {Infeld},\ and\ \citenamefont
  {Hoffmann}}]{Einstein:1938yz}%
  \BibitemOpen
  \bibfield  {author} {\bibinfo {author} {\bibfnamefont {A.}~\bibnamefont
  {Einstein}}, \bibinfo {author} {\bibfnamefont {L.}~\bibnamefont {Infeld}}, \
  and\ \bibinfo {author} {\bibfnamefont {B.}~\bibnamefont {Hoffmann}},\ }\href
  {\doibase 10.2307/1968714} {\bibfield  {journal} {\bibinfo  {journal} {Annals
  Math.}\ }\textbf {\bibinfo {volume} {39}},\ \bibinfo {pages} {65} (\bibinfo
  {year} {1938})}\BibitemShut {NoStop}%
\bibitem [{\citenamefont {Poisson}\ and\ \citenamefont {Will}(2014)}]{Gravity}%
  \BibitemOpen
  \bibfield  {author} {\bibinfo {author} {\bibfnamefont {E.}~\bibnamefont
  {Poisson}}\ and\ \bibinfo {author} {\bibfnamefont {C.~M.}\ \bibnamefont
  {Will}},\ }\href@noop {} {\emph {\bibinfo {title} {{Gravity}}}}\ (\bibinfo
  {publisher} {Cambridge, UK: Cambridge University Press},\ \bibinfo {year}
  {2014})\BibitemShut {NoStop}%
\bibitem [{\citenamefont {Blandford}\ and\ \citenamefont
  {Teukolsky}(1976)}]{BT}%
  \BibitemOpen
  \bibfield  {author} {\bibinfo {author} {\bibfnamefont {R.}~\bibnamefont
  {Blandford}}\ and\ \bibinfo {author} {\bibfnamefont {S.~A.}\ \bibnamefont
  {Teukolsky}},\ }\href {\doibase 10.1086/154315} {\bibfield  {journal}
  {\bibinfo  {journal} {Astrophys. J.}\ }\textbf {\bibinfo {volume} {205}},\
  \bibinfo {pages} {580} (\bibinfo {year} {1976})}\BibitemShut {NoStop}%
\bibitem [{\citenamefont {Damour}\ and\ \citenamefont
  {Taylor}(1992)}]{Damour:1991rd}%
  \BibitemOpen
  \bibfield  {author} {\bibinfo {author} {\bibfnamefont {T.}~\bibnamefont
  {Damour}}\ and\ \bibinfo {author} {\bibfnamefont {J.~H.}\ \bibnamefont
  {Taylor}},\ }\href {\doibase 10.1103/PhysRevD.45.1840} {\bibfield  {journal}
  {\bibinfo  {journal} {{Phys. Rev. D}}\ }\textbf {\bibinfo {volume} {45}},\
  \bibinfo {pages} {1840} (\bibinfo {year} {1992})}\BibitemShut {NoStop}%
\bibitem [{\citenamefont {Wex}\ and\ \citenamefont
  {Kopeikin}(1999)}]{Wex:1998wt}%
  \BibitemOpen
  \bibfield  {author} {\bibinfo {author} {\bibfnamefont {N.}~\bibnamefont
  {Wex}}\ and\ \bibinfo {author} {\bibfnamefont {S.}~\bibnamefont {Kopeikin}},\
  }\href {\doibase 10.1086/306933} {\bibfield  {journal} {\bibinfo  {journal}
  {Astrophys. J.}\ }\textbf {\bibinfo {volume} {514}},\ \bibinfo {pages} {388}
  (\bibinfo {year} {1999})},\ \Eprint {http://arxiv.org/abs/astro-ph/9811052}
  {arXiv:astro-ph/9811052} \BibitemShut {NoStop}%
\bibitem [{\citenamefont {Sellentin}\ \emph {et~al.}(2014)\citenamefont
  {Sellentin}, \citenamefont {Quartin},\ and\ \citenamefont
  {Amendola}}]{Sellentin:2014zta}%
  \BibitemOpen
  \bibfield  {author} {\bibinfo {author} {\bibfnamefont {E.}~\bibnamefont
  {Sellentin}}, \bibinfo {author} {\bibfnamefont {M.}~\bibnamefont {Quartin}},
  \ and\ \bibinfo {author} {\bibfnamefont {L.}~\bibnamefont {Amendola}},\
  }\href {\doibase 10.1093/mnras/stu689} {\bibfield  {journal} {\bibinfo
  {journal} {Mon. Not. Roy. Astron. Soc.}\ }\textbf {\bibinfo {volume} {441}},\
  \bibinfo {pages} {1831} (\bibinfo {year} {2014})},\ \Eprint
  {http://arxiv.org/abs/1401.6892} {arXiv:1401.6892 [astro-ph.CO]} \BibitemShut
  {NoStop}%
\bibitem [{\citenamefont {Wang}\ \emph {et~al.}(2022)\citenamefont {Wang},
  \citenamefont {Liu}, \citenamefont {Zhao},\ and\ \citenamefont
  {Shao}}]{Wang:2022kia}%
  \BibitemOpen
  \bibfield  {author} {\bibinfo {author} {\bibfnamefont {Z.}~\bibnamefont
  {Wang}}, \bibinfo {author} {\bibfnamefont {C.}~\bibnamefont {Liu}}, \bibinfo
  {author} {\bibfnamefont {J.}~\bibnamefont {Zhao}}, \ and\ \bibinfo {author}
  {\bibfnamefont {L.}~\bibnamefont {Shao}},\ }\href {\doibase
  10.3847/1538-4357/ac6b99} {\bibfield  {journal} {\bibinfo  {journal}
  {Astrophys. J.}\ }\textbf {\bibinfo {volume} {932}},\ \bibinfo {pages} {102}
  (\bibinfo {year} {2022})},\ \Eprint {http://arxiv.org/abs/2203.02670}
  {arXiv:2203.02670 [gr-qc]} \BibitemShut {NoStop}%
\bibitem [{\citenamefont {Hastings}(1970)}]{Hastings:1970aa}%
  \BibitemOpen
  \bibfield  {author} {\bibinfo {author} {\bibfnamefont {W.~K.}\ \bibnamefont
  {Hastings}},\ }\href {\doibase 10.1093/biomet/57.1.97} {\bibfield  {journal}
  {\bibinfo  {journal} {Biometrika}\ }\textbf {\bibinfo {volume} {57}},\
  \bibinfo {pages} {97} (\bibinfo {year} {1970})}\BibitemShut {NoStop}%
\bibitem [{\citenamefont {Skilling}(2004)}]{Skilling:2004}%
  \BibitemOpen
  \bibfield  {author} {\bibinfo {author} {\bibfnamefont {J.}~\bibnamefont
  {Skilling}},\ }\href {\doibase 10.1063/1.1835238} {\bibfield  {journal}
  {\bibinfo  {journal} {AIP Conference Proceedings}\ }\textbf {\bibinfo
  {volume} {735}},\ \bibinfo {pages} {395} (\bibinfo {year}
  {2004})}\BibitemShut {NoStop}%
\bibitem [{\citenamefont {Skilling}(2006)}]{Skilling:2006gxv}%
  \BibitemOpen
  \bibfield  {author} {\bibinfo {author} {\bibfnamefont {J.}~\bibnamefont
  {Skilling}},\ }\href {\doibase 10.1214/06-BA127} {\bibfield  {journal}
  {\bibinfo  {journal} {Bayesian Analysis}\ }\textbf {\bibinfo {volume} {1}},\
  \bibinfo {pages} {833} (\bibinfo {year} {2006})}\BibitemShut {NoStop}%
\bibitem [{\citenamefont {Liu}\ \emph {et~al.}(2011)\citenamefont {Liu},
  \citenamefont {Verbiest}, \citenamefont {Kramer}, \citenamefont {Stappers},
  \citenamefont {van Straten},\ and\ \citenamefont {Cordes}}]{Liu:2011cka}%
  \BibitemOpen
  \bibfield  {author} {\bibinfo {author} {\bibfnamefont {K.}~\bibnamefont
  {Liu}}, \bibinfo {author} {\bibfnamefont {J.~P.~W.}\ \bibnamefont
  {Verbiest}}, \bibinfo {author} {\bibfnamefont {M.}~\bibnamefont {Kramer}},
  \bibinfo {author} {\bibfnamefont {B.~W.}\ \bibnamefont {Stappers}}, \bibinfo
  {author} {\bibfnamefont {W.}~\bibnamefont {van Straten}}, \ and\ \bibinfo
  {author} {\bibfnamefont {J.~M.}\ \bibnamefont {Cordes}},\ }\href {\doibase
  10.1111/j.1365-2966.2011.19452.x} {\bibfield  {journal} {\bibinfo  {journal}
  {Mon. Not. Roy. Astron. Soc.}\ }\textbf {\bibinfo {volume} {417}},\ \bibinfo
  {pages} {2916} (\bibinfo {year} {2011})},\ \Eprint
  {http://arxiv.org/abs/1107.3086} {arXiv:1107.3086 [astro-ph.HE]} \BibitemShut
  {NoStop}%
\bibitem [{\citenamefont {{Foreman-Mackey}}\ \emph {et~al.}(2013)\citenamefont
  {{Foreman-Mackey}}, \citenamefont {{Hogg}}, \citenamefont {{Lang}},\ and\
  \citenamefont {{Goodman}}}]{2013PASP..125..306F}%
  \BibitemOpen
  \bibfield  {author} {\bibinfo {author} {\bibfnamefont {D.}~\bibnamefont
  {{Foreman-Mackey}}}, \bibinfo {author} {\bibfnamefont {D.~W.}\ \bibnamefont
  {{Hogg}}}, \bibinfo {author} {\bibfnamefont {D.}~\bibnamefont {{Lang}}}, \
  and\ \bibinfo {author} {\bibfnamefont {J.}~\bibnamefont {{Goodman}}},\ }\href
  {\doibase 10.1086/670067} {\bibfield  {journal} {\bibinfo  {journal} {{Publ.
  Astron. Soc. Pac.}}\ }\textbf {\bibinfo {volume} {125}},\ \bibinfo {pages}
  {306} (\bibinfo {year} {2013})},\ \Eprint {http://arxiv.org/abs/1202.3665}
  {arXiv:1202.3665 [astro-ph.IM]} \BibitemShut {NoStop}%
\bibitem [{\citenamefont {Foreman-Mackey}(2016)}]{corner}%
  \BibitemOpen
  \bibfield  {author} {\bibinfo {author} {\bibfnamefont {D.}~\bibnamefont
  {Foreman-Mackey}},\ }\href {\doibase 10.21105/joss.00024} {\bibfield
  {journal} {\bibinfo  {journal} {The Journal of Open Source Software}\
  }\textbf {\bibinfo {volume} {1}},\ \bibinfo {pages} {24} (\bibinfo {year}
  {2016})}\BibitemShut {NoStop}%
\bibitem [{\citenamefont {Smits}\ \emph {et~al.}(2009)\citenamefont {Smits},
  \citenamefont {Kramer}, \citenamefont {Stappers}, \citenamefont {Lorimer},
  \citenamefont {Cordes},\ and\ \citenamefont {Faulkner}}]{Smits:2008cf}%
  \BibitemOpen
  \bibfield  {author} {\bibinfo {author} {\bibfnamefont {R.}~\bibnamefont
  {Smits}}, \bibinfo {author} {\bibfnamefont {M.}~\bibnamefont {Kramer}},
  \bibinfo {author} {\bibfnamefont {B.}~\bibnamefont {Stappers}}, \bibinfo
  {author} {\bibfnamefont {D.~R.}\ \bibnamefont {Lorimer}}, \bibinfo {author}
  {\bibfnamefont {J.}~\bibnamefont {Cordes}}, \ and\ \bibinfo {author}
  {\bibfnamefont {A.}~\bibnamefont {Faulkner}},\ }\href {\doibase
  10.1051/0004-6361:200810383} {\bibfield  {journal} {\bibinfo  {journal}
  {Astron. Astrophys.}\ }\textbf {\bibinfo {volume} {493}},\ \bibinfo {pages}
  {1161} (\bibinfo {year} {2009})},\ \Eprint {http://arxiv.org/abs/0811.0211}
  {arXiv:0811.0211 [astro-ph]} \BibitemShut {NoStop}%
\end{thebibliography}%

\end{document}